\documentclass[12pt]{article}

\usepackage[paper=portrait,pagesize]{typearea}

\usepackage{amsmath}

\usepackage{graphicx} 

\usepackage{appendix}

\usepackage{setspace}
\usepackage{algorithm}
\usepackage{algpseudocode}

\usepackage{longtable}
\usepackage{pdflscape}
\usepackage[table]{xcolor}
\usepackage{booktabs}

\usepackage{footmisc} 

\usepackage[utf8]{inputenc}
\usepackage[authordate, backend=biber]{biblatex-chicago}
\DeclareBibliographyDriver{data}{%
  \printnames{author}%
  \setunit{\labelnamepunct}%
  \printfield{title}%
  \setunit{\addcomma\space}%
  \printfield{year}%
  \finentry
}

\title{Generative AI as a Safety Net for Survey Question Refinement\thanks{We gratefully acknowledge FORMAS (2016-00228, PI: Ellen Lust) and The Swedish Research Council (2016-01687, PI: Ellen Lust, and E0003801, PI: Pam Fredman) for financial support; Megan Baxter for data collection support; Emmanuel Kwabena Asare and Kiara Castaman for qualitative coding support; Rose Shaber-Twedt for editing support; Ellen Lust for insights and feedback; and the MiMac Conference for early encouragement and feedback.}}
\author{Erica Ann Metheney\thanks{Researcher/Statistician/Head of Data, Governance and Local Development Institute, University of Gothenburg},
Lauren Yehle\thanks{PhD Student, University of Gothenburg}}
\date{September 2025}

\begin{document}

\maketitle

\clearpage

\begin{abstract}
\singlespacing 
\small         
Writing survey questions that easily and accurately convey their intent to a variety of respondents is a demanding and high-stakes task. Despite the extensive literature on best practices, the number of considerations to keep in mind is vast and even small errors can render collected data unusable for its intended purpose. The process of drafting initial questions, checking for known sources of error, and developing solutions to those problems requires considerable time, expertise, and financial resources. Given the rising costs of survey implementation and the critical role that polls play in media, policymaking, and research, it is vital that we utilize all available tools to protect the integrity of survey data and the financial investments made to obtain it. Since its launch in 2022, ChatGPT and other generative AI model platforms have been integrated into everyday life processes and workflows, particularly pertaining to text revision. While many researchers have begun exploring how generative AI may assist with questionnaire design, we have implemented a prompt experiment to systematically test what kind of feedback on survey questions an average ChatGPT user can expect. Results from our zero--shot prompt experiment, which randomized the version of ChatGPT and the persona given to the model, shows that generative AI is a valuable tool today, even for an average AI user, and suggests that AI will play an increasingly prominent role in the evolution of survey development best practices as precise tools are developed.

\end{abstract}

\clearpage

\section{Introduction}

Whether you are a social scientist measuring public opinion, a doctor assessing a patient’s health status, a government conducting a census, or a corporation trying to understand customer preferences, survey data can provide information related to your most pressing questions. This is because a survey is simply a structured questionnaire that ``...consists of a set of standardized questions with a fixed scheme, which specifies the exact wording and order of the questions, for gathering information from respondents" \parencite{cheung_structured_2014}. This generic data collection format is an incredibly adaptable tool that generates data that, when paired with statistical methods, can produce incredible insights\parencite{payne1951}. 

However, the creation of valid and reliable survey questionnaires that produce high-quality data is a laborious task that requires extensive time, financing, and expertise. This is partly because survey designers not only have to worry about foundational semantic difficulties of survey questions, i.e., problems affiliated with meaning, complexity, and concepts, but also task difficulties, i.e., problems affiliated with recall, reporting, readability, and recording \parencite{presser1994survey}. These difficulties are often in direct competition; for example, a researcher must decide whether it is better to use a shorter question that is faster to read or a longer version that might provide more clarity. Fortunately, the widespread use of surveys has resulted in a vast array of rigorously tested tools, guidelines, and best practices for survey development that can be used to reveal problems that even the most experienced team or individual may not be able to identify on their own.

Although there is extensive research on how one should design a survey and even tools are available to facilitate the implementation of best practices, practical implementation of all best practices is limited.  Tools like the Question Appraisal System (QAS) that helps identify potential problems by offering over 25 checks per item are time consuming \parencite{willis1999question}, focus groups and cognitive interviews that dig into a respondent's thought process when answering each question are expensive and require the time of trained experts to administer them \parencite{fowler_improving_1995}, and the sheer number of guidelines and recommendations for word choice, length, order of questions and order of responses are so numerous that even the most trained and careful individual or team can easily let things slip through the cracks \parencite{converse_survey_1986}.

One can easily see a multitude of scenarios where best practices cannot be followed practically, confidently, or wholistically -- a student working on their thesis, a rapid response team needing to gather data after a crisis, or an underfunded research group. Indeed, the state-of-the-art in survey design follows the total survey error (TSE) framework, which calls for balancing data quality with the reality of resource constraints (time and budgetary) \parencite{biemer2010total}. The TSE framework is broadly structured into two categories: measurement - how we measure, and representation - who/what we measure. When data quality is intentionally compromised within the TSE framework, concessions are usually made within the representation side due to sampling's direct impact on total cost and timelines. However, resource constraints have an equally pressing, though less direct and often less intentional, impact on the measurement side. As the cost of collecting survey data rises and response rates fall \parencite{wagner_using_2020}, it is more crucial than ever that fielded questionnaires are of the highest quality.

Since the release of ChatGPT--3.5 in November of 2022, researchers have been exploring and testing how generative AI can improve and expand current best practices across all disciplines. Of the working and published studies regarding generative AI, most address its applicability to fields like education, law, and computer programming rather than research applications for social science \parencite{liu2023summary,wu2023brief}. However, in recent years the prevalence of AI in social science has been growing, amid concerns of bias and other ethical issues \parencite{bail_can_2024}.

When considering how to integrate AI into survey methodology, and recognizing the work currently being done, one sees two distinct research paths: 1) developing new tools that expand the array of best practices, opening doors that did not previously exist, and 2) using AI to make existing best practices more efficient, effective, and accessible. Scholars working in Path 1 are seeking to leverage AI to its greatest extent. However, much of that scholarship focuses on niche applications of survey design \parencite{paduraru2024adaptive} and is restricted to digital survey modes \parencite{esra2025, sturgissocbot}. Furthermore, the tools resulting from this path will likely require access to extensive computing resources or costly proprietary software, raising accessibility limitations. Simultaneously, other scholars have begun investigating Path 2 \parencite{chatGPTest}. However, much of the work in this area is example-based or is restricted to a specific case-study \parencite{fichtner2024ex}. 

\begin{figure}[htbp]
\centering
\includegraphics[width=0.85\textwidth]{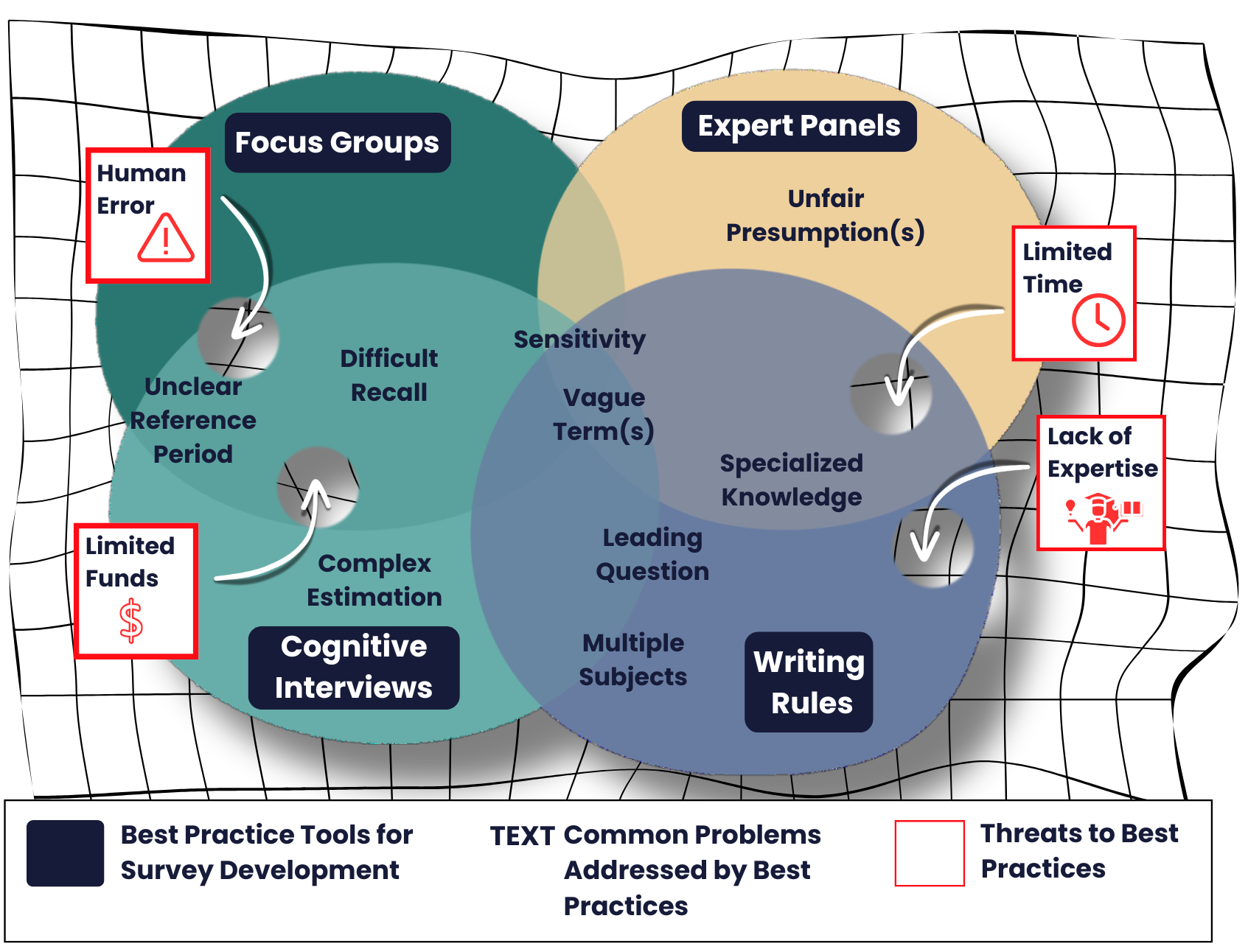}
\caption{Illustration depicting how we can conceptualize AI as a safety net for survey methodology best practices based on research 
The appendix elaborates on highly effective best practices, such as cognitive interviews, writing rules, and expert panels, but resource constraints reduce the ability to use these tools well or even at all. 
}
\label{fig:safetynet}
\end{figure}

Our work falls into Path 2; however, we utilize a systematic, pre-registered experimental design to understand:  ``How can we use generative AI \textit{today} to supplement the best practices we already have?", ``What kind of feedback can an average ChatGPT user expect to receive regarding their survey questions?", and ``How does choice of model and persona impact the output?".  Thus, as shown in Figure \ref{fig:safetynet}, we seek to learn how we can best use AI as a safety net when our best practices fail or are simply unavailable due to a lack of time, money, or expertise.

Our $2\times3$ factorial experiment aims to understand what kind of feedback an average AI--user could expect to get on their survey questions. Our experiment consists of 6 unique treatments that randomize the version of GPT  used (3.5 and 4.0)\footnote{These were the available models during data collection from September to December 2023.} and the persona (none, survey design expert, linguist) given to the AI model in the prompt. The prompts were kept short and simple, to reflect the type of prompts an average user might write. An example prompt reads:

\begin{quote}
\textit{You are a survey design expert who drafted a survey in English that will be administered to respondents from a single cultural background that speak English as their first language. For the following question, give me up to 5 features of the question that could cause the respondents to interpret the question differently. Do you own a car?}
\end{quote}

Each treatment was applied to 262 questions obtained from three sources: the Gallup Q12 Survey, the World Values Survey (WVS), and the Local Governance Performance Index survey (LGPI). The content of the responses from the AI model was qualitatively coded using a codebook that summarizes common survey question problems identified in the survey methodology literature. 

\begin{table}[htbp]
\centering
\caption{Qualitative Codes} \label{tab:codebook_simple}
\begin{tabular}[t]{|c|p{15em}|}
\hline
\textbf{Code} & \textbf{Label} \\ \hline

1     & Vague term or phrase \\ \hline
2     & Specialized knowledge \\ \hline
3     & Syntax problem \\ \hline
4     & Unfair presumption \\ \hline
5     & Double barreled \\ \hline
6     & Undefined/ill-defined reference period \\ \hline
7     & Difficult recall or reference period \\ \hline
8     & Complex estimation/computation \\ \hline
9     & Sensitivity \\ \hline
10    & Leading/biasing question \\ \hline
11    & Answer set \\ \hline
NOTA  & None of the above \\ \hline
SysVar & Systematic Variation (subset of NOTA) \\ \hline

\end{tabular}
\end{table}

Table \ref{tab:codebook_simple} summarizes the codes while the full codebook can be found in Appendix \ref{Appendix:Qual}. Each treatment--question pair resulted in 5 statements whose content was individually coded. We then created a new set of indicator variables at the treatment--question level, one for each code, that indicates if the code appears in any of the 5 statements.

\section{Methodology}
\subsection{Experimental Design}
We use a 2$\times$3 factorial design resulting in 6 unique treatments. The two factors we randomize are the version of GPT (3.5 and 4.0) and the persona (none, Survey Design Expert, Linguist) given to the AI model. 

New generative AI models and updated versions of existing models continue to be released at a rapid pace. While the commonly held assumption is that ``new is better," we randomize the version of ChatGPT  to systematically test if this holds across all aspects of the evaluation task. It also provides empirical evidence for the importance of re-evaluating the performance of AI-based tools and strategies as new versions and models are released. 

Prompts given to LLMs enforce rules, automate processes, and ensure specific qualities and quantities of generated output; they also helps the researcher reflect the researcher's specific desires \parencite{white_prompt_2023}. Our control condition for persona provides a baseline of what the AI can generate with fewer restrictions. Then we specify the type of persona as (1) a survey design expert or (2) a linguist. Both types of expertise can inform survey design and might alter the types of problems flagged. 

The general format of our instruction prompt is: 
\begin{quote}
\textit{\{Persona\} drafted a survey in English that will be administered to respondents from a single cultural background that speak English as their first language. For the following question, give me up to 5 features of the question that could cause the respondents to interpret the question differently. \{Question\}}
\end{quote}

\noindent The prompt includes the phrase ``a survey in English that will be administered to respondents from a single cultural background that speak English as their first language" to fix the cultural-linguistic context. This was an attempt to reduce the likelihood that the model provides comments on problems regarding language or cultural variation. 

The decision to ask for 5 features was made arbitrarily to balance the likelihood of getting all possible feedback from the AI against the possibility of the output becoming too large and to provide consistent structure to the format of the AI's responses. The distribution of the number of codes per treatment-question pair suggests that 5 was a sufficient limit. 

Lastly, our prompt asks the AI to determine ``features of the question that could cause the respondents to interpret the question differently." Instead of providing a codebook for a specific assessment tool, such as the QAS--99, we wanted to ask the AI to evaluate the heart of the issue in survey design, do all respondents interpret the question the same way--in hopes of obtaining the most comprehensive range of feedback one can expect from the models. 

This design and the following hypotheses were pre-registered on OSF on November 23, 2023: \url{https://osf.io/4q26k}.

\begin{itemize}
    \item  H1 (Main Effect of Model): We expect that version 4 of GPT is more likely to flag all codes than version 3.5. 
    \item H2 (Effect of persona compared to unspecified): Providing a specific persona (survey design expert or linguist) will be more likely to flag all codes than when no specific persona is given. 
    \item H3 (Effect of Linguist Persona on Specific Codes): Providing the model with the persona linguist will increase the likelihood of flagging Code 1: Vague term or phrase and Code 3: Syntax compared to a when the persona is unspecified or assigned as survey design expert.
\end{itemize}

\subsection{Power Analysis}
To determine a sufficient sample size, we performed a power analysis via simulation using the \texttt{powerSim} function in the \texttt{simr} package \parencite{simrPackage} in R  \parencite{Rsoftware} to understand our power to detect main and interaction effects in our multilevel logistic model \parencite{arend2019statistical}. While we also plan to run ordinary multilevel models (MLM), the sample size needs for a logistic MLM are greater, so we only run a power analysis for the logistic MLM \parencite{ali_sample_2019, moineddin_simulation_2007}.  

We considered sample sizes (number of questions) ranging from 200 to 350, direct effect sizes of 0.1, 0.41, and 0.7 (corresponding generally to small, medium, and large effects), and interaction effects of 0.05, 0.1, and 0.2. Given this is the first study of its kind, we had no previous work on which we could base our choice of effects sizes. Because each effect will be tested for 11 codes, we also run the power analysis with a Bonferroni correction for multiple testing ($\alpha$ = 0.05/11) and without ($\alpha$ = 0.05). While in practice one would use the Holm correction, for simplicity we use the more conservative Bonferroni correction in the simulation. 

\subsection{Sourcing Experimental Units}
We sourced 262 questions from surveys developed by Gallup, the World Values Survey Association (WVS), and the Governance and Local Development Institute (GLD). Specific questions from Gallup, WVS, and the LGPI were selected based on their ability to be posed as ``stand alone" questions. Questions were excluded if they relied on answer set structure, survey logic, were not phrased as a question, or were redundant. Questions that were structured as fill in the blank (for example, ``I am...with my government.") were also excluded.\footnote{Additional details can be found in Appendix \ref{Appendix:SourcingEUs}.} 

From Gallup, we selected the publicly available Q12 survey, containing 13 questions to measure employee engagement \parencite{gallup_overview}. We selected questions from the 7th wave of WVS. The WVS has been translated and implemented in over 90 countries since the first wave in 1981 \parencite{wvs_methodology}. Finally, the 2019 version of the LGPI survey conducted by GLD, was implemented in 3 countries, requiring six translations \parencite{gld_overview}. We believe these three sources provide a range of refinement and topics. We expect the AI to identify the most potential  issues for the LGPI because this survey covers the largest range of topics with the fewest refinement iterations. GLD is a newer organization than Gallup and WVS, with fewer opportunities to refine errors in the questions. Thus, we anticipate fewer problems to be flagged by the AI for the WVS questions because WVS had six prior waves of experience  and covers fewer topics. Finally, the short, precise Q12 survey from Gallup is the mostly widely used and vetted and thus should have the fewest unresolved problems for the AI to identify.

\subsection{Obtaining GPT Output}
Each survey question in our sample serves as an experimental unit (EU) that independently receives all six treatments. We used an R script to create all experimental and formatting prompts. Procedure \ref{procedure_summary1} describes the process of obtaining the AI output. The final step was implemented using a custom R script. All code is available in the Harvard Dataverse \parencite{DVRepo}. 

\begin{algorithm}
\caption{Steps to Generate Experimental Output}\label{procedure_summary1}
\begin{algorithmic}
\State $n =$ number of survey questions in the sample
\State $t =$ number of unique treatments
\For{$treatment\gets 1, t$}
\For{$question\gets 1, n$}
\State Start a new GPT Chat on \url{chapgpt.com}
\State Enter the prompt as defined by the $treatment$
\State Enter the $question$
\State Enter the formatting prompt
\EndFor
\EndFor
\State Export all chats to a .json file
\State Convert .json to .xlsx file
\end{algorithmic}
\end{algorithm}

By starting a new chat before administering every prompt, we ensured independence of the AI's responses and prevented it from learning across treatments\footnote{The current memory feature of ChatGPT was not available when the data was collected from September to December 2023.}. We utilized the web-version of ChatGPT with default parameter values to mimic the experience that a typical researcher seeking assistance from ChatGPT might encounter. 

\subsection{Qualitative Coding}
The study's goal is to flag problems from the literature, so researchers can rewrite the question before pretesting and implementation. Thus, we need a coding scheme that focuses on \textit{fixable} problems like word choice, syntax, ambiguity, etc. The best survey questionnaire assessment tool is the Question Appraisal System with over 25 questions to identify issues with a specific research question \parencite{willis1999question}. We adopt a truncated version of this Appraisal System with fewer codes based on an array of sources \parencite{fowler2011, presser1994survey, bais_can_2019, fowler2014problem, bais_can_2019, holbrook2006impact, ongena2006methods, van2004studying}. We also adapt Rothgeb et al.'s scheme originally utilized to compare pre-testing techniques' abilities to identify problems\parencite{rothgeb2007} . This focuses on four types of problems: question content, question structure, retrieval from memory, and judgment/evaluation, resulting in 10 subcodes and 2 emergent codes. Unlike other schemes, this does not rely on respondent behavior/answers and provides the most specificity about word choice and question content. If the statement did not qualify as one of the 11 codes, it was coded as None of the Above. 

\subsection{Analysis}

Given the factorial experimental design and that every experimental unit receives each treatment, we use 2-level hierarchical regression models to test our pre--registered hypotheses. When the outcome is binary, we fit the following model: 
\begin{align}
logit(y_{ij})  = &  \beta_0 + \beta_M\mathbf{1}_{GPT-4} +  \beta_{T2}\mathbf{1}_{Survey Design Expert} \\
& + \beta_{T3}\mathbf{1}_{Linguist} + u_i + \varepsilon_{ij} \nonumber
\end{align}
where $y_{ij}$ takes on the value 0 or 1, indicating if a specific code was flagged for the output from question $i$ after receiving treatment $j$. The $\mathbf{1}$s represent indicator variables for the noted factor level and the $\beta$s are the corresponding coefficients. The term $u_i$ is the random intercept associated with the question and $\varepsilon_{ij}$ represents the error. The inclusion of the random question intercept addresses the fact that each treatment was applied to every question.

 We will also fit 2-level linear probability models for each binary outcome:
\begin{align}
y_{ij} =  & \beta_0 + \beta_M\mathbf{1}_{GPT-4} +  \beta_{L1}\mathbf{1}_{Survey Design Expert}\\ 
& + \beta_{L2}\mathbf{1}_{Linguist} + u_i + \varepsilon_{ij}. \nonumber
\end{align}
When the results are robust for the multilevel logistic and multilevel linear probability models, we present the multilevel linear probability models, due to their ease of interpretation.

When assessing the effect of model and persona on the likelihood of generating content related to each of the 11 primary codes, we will assess statistical significance using $p$-values that have been adjusted using the Holm's correction for multiple testing. All other regression models and statistical tests will assess significance at $\alpha = 0.05$. 

Additionally, we present a series of exploratory analyses, some of which were suggested as such in the pre--analysis plan and others that emerged after qualitative analysis. First, we analyze the effect of question source on the total number of codes and likelihood of flagging individual codes, using modified versions of the 2-level hierarchical models. We also explore if there is any structure or information contained in the placement of the flagged codes, i.e., are certain codes more likely to appear earlier or later in the 5 statements? We run a one-way ANOVA test to determine if all codes have the same average statement placement. Should we find a statistically significant result, we run a Tukey HSD test to determine which codes have similar and different average placement.

\section{Results}

We find partial support for all three preregistered hypotheses and no statistically significant evidence to the contrary.\footnote{The hypotheses and analysis plan were submitted on OSF on November 23, 2023 \url{https://osf.io/4q26k}.} Approximately 81\% of the individual statements contained content related to a single code, 3\% related to two codes, and 15\% provided no qualitatively codeable content. On average, across the 5 statements at the question--treatment level, there were 2.69 qualitative codes. Notably,  Code 1 (vague term) was coded in 98\% of question-treatment sets due to the code being too broadly defined. Therefore, the results cannot be meaningfully investigated, but we present the results without interpretation for completeness. We investigate the effect of model and persona on the total number of codes flagged at the question--treatment level and the likelihood of flagging a specific code at the question--treatment level. Statistical significance of the regression results is determined based on a Holm's multiple testing correction. All other statistical tests are assessed at $\alpha = 0.05$.

\subsection{Model Effect} 
\begin{figure}[htbp]
    \centering
    \includegraphics[width=0.75\textwidth]{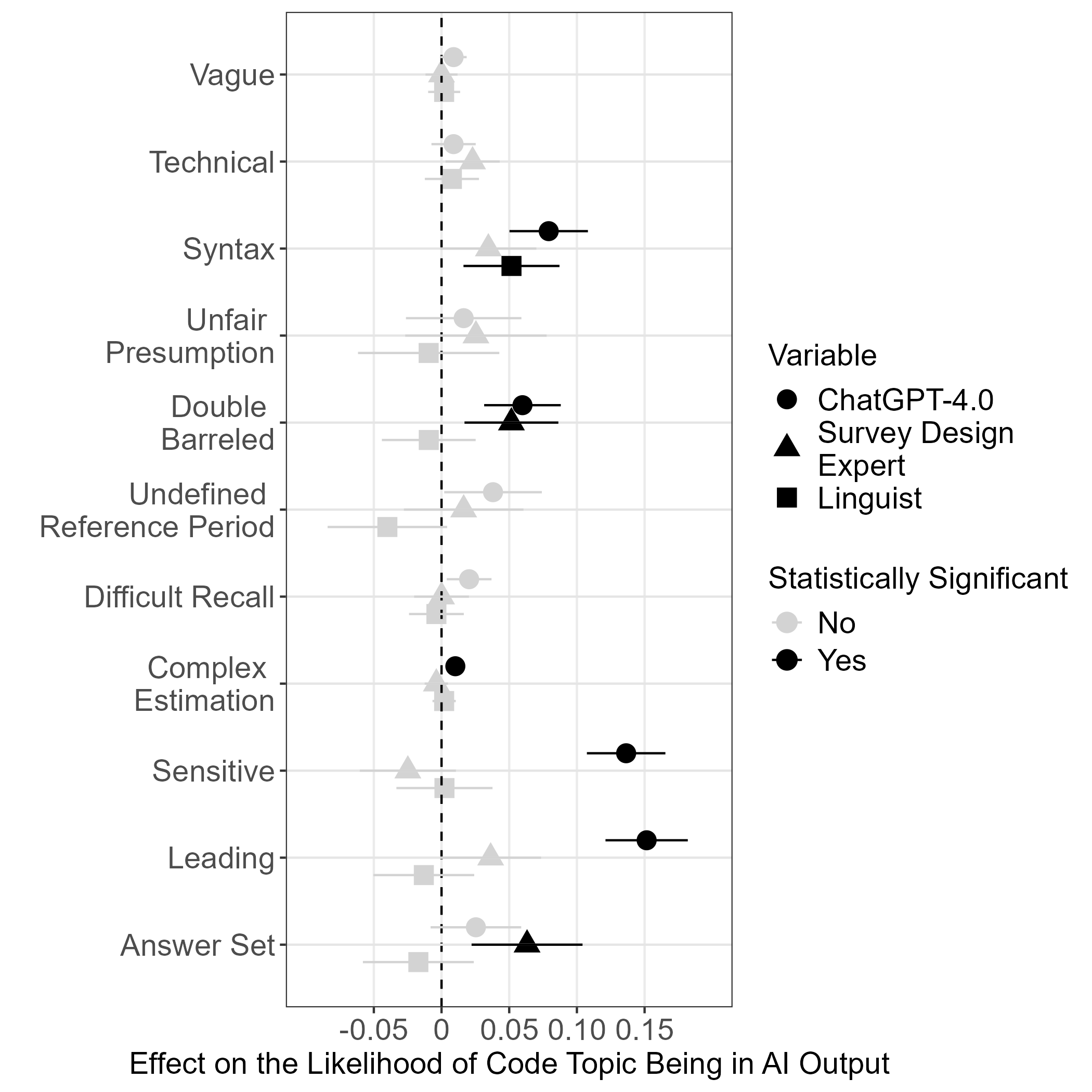}
    \caption{Marginal Effects Plot Showing the Impact of Each Experimental Attribute on the Likelihood of Each Code Topic Appearing in the  AI Output. Statistical significance determined using $p$-values adjusted for multiple testing with Holm's correction.}
    \label{fig:expeffects}
\end{figure}

GPT--3.5 and GPT--4.0 content systematically produce different content, as indicated by the circle icons in Figure \ref{fig:expeffects}. Compared to the earlier model, GPT--3.5, GPT--4.0 produces, on average 0.55 ($p<$0.001), more codes and is more likely to produce Code 3--Syntax problem(\(\beta = 0.08\)), Code 5--Double barreled (\(\beta = 0.06\)), and Code 8--Complex estimatation (\(\beta = 0.01\)), and, with the largest effect sizes of the study, Code 9--Sensitivity (\(\beta = 0.14\)) and Code 10--Leading/biasing (\(\beta = 0.15\))\footnote{See the tables in Appendix \ref{Appendix:FullResults} for full results.}. Finally, GPT--4.0 is 17 percentage points ($pp$) less likely to produce NOTA (none of the above) statements. Notably, the more advanced GPT--4.0 model returns more content related to codes that require more consideration of what the respondent may think or feel (e.g., sensitivity or leading) beyond what is directly stated in the question (e.g., missing recall period or syntax). 

\subsection{Persona Effect}
As expected \parencite{white_prompt_2023}, prompting the AI with either persona yielded substantially different outputs than without a persona. The Survey Design Expert persona produces 0.22 ($p<0.001$) more codes on average, is 8$pp$ less likely to produce NOTA ($p< 0.01$), and more frequently produces  Code 5--Double barreled ($\beta = 0.05, p<0.05$) and Code 11--Answer set ($\beta = 0.06, p<0.05$) compared to no persona. Interestingly, these are two of the codes most uniquely associated with survey methodology in the codebook. Meanwhile, the Linguist persona is only more likely to produce Code 3--Syntax (\(\beta = 0.05\)), the code most closely related to the expertise of a linguist. These results highlight the importance of strategic persona selection when seeking feedback on survey questions.

\subsection{Unpacking None of the Above}
NOTA was the second most frequent code, applied to nearly 17\% of all question-treatment pairs. NOTA-coded statements include a range of content from meaningless output to helpful but off-task feedback. We denoted one version of the latter that occurred at high frequency (41\% of all NOTA), Systematic Variation (SysVar). In SysVar cases, instead of explaining why respondents might \textit{interpret} the question differently, the AI explained why the respondents might \textit{answer} differently. While this is distinctly off-task, the information can be viewed as suggestions for future analyses or control variables that should be included in the survey. For example, 

\begin{quote}
    \textit{...someone who works in a tech-related field might have a different perspective than someone in a field less influenced by modern technology. Furthermore, historical events, media coverage, or influential figures in the community can shape perceptions about the impact of science and technology.}
\end{quote}

\noindent From this output, one could consider the value of adding an employment sector question to their survey to control for an additional source of variation in their analysis. We find that this type of SysVar content was 5$pp$ ($p<0.05$) less likely to be generated by GPT--4.0 compared to GPT--3.5 and by the Survey Design Expert compared to no persona. 

Also outside the scope of the instructions, the AI offered feedback relevant to context ``\textit{...if the survey is part of a larger questionnaire about legal documentation or citizenship, respondents might interpret the question differently than if its part of a survey about child health or education,}" and question order/priming ``\textit{Without the context of the entire survey, its possible that other questions in the survey address related or overlapping topics, which could affect how a respondent interprets this particular question. For instance, if there's another question about transportation means or assets in general, it might cause some confusion or affect how respondents approach this question.}" Qualitatively reading output like these examples emphasizes the importance of reviewing content for novel material.

\subsection{Validity Checks}
We performed three validity checks to access the quality of the AI output. First, we examined if the AI produced codes as one would expect, given the nature of the question source. Second, we examined the structure of the output by performing a statement order analysis. Third, we compared the AI--generated codes to ones produced by a human. 

\subsubsection{Source of the Question}
We find that ChatGPT often flagged issues in expected ways depending on the question's source. We consider the Gallup Q12 survey questions of employee engagement the most refined source, given its administration to over 3.3 million workers across 100,000+ workplace teams \parencite{gallup_overview}. The other two sources are academic rather than corporate, covering a wider range of topics and with fewer implementations than the Gallup Q12--three prior waves for LGPI \parencite{lgpi_overview} and six for WVS \parencite{wvs_overview}. Pursuant to this logic, our results indicate that Gallup-sourced questions result in 0.31 fewer codes on average, ($p<0.01$), at the statement-treatment level compared to the LGPI. Conversely, we also find that WVS-sourced questions, which have undergone twice as many rounds of implementation compared to the LGPI, yield 0.19 ($p<0.001$) \textit{more} codes.

When looking at the likelihood of specific codes occurring, we find no evidence that Gallup-sourced questions are more or less likely to return any of the 10 codes compared to the LGPI. Alternatively, we find that the WVS-sourced questions are more likely to return Code 2--Technical Knowledge ($\beta = 0.07, p\le 0.01$), a sensible finding given that the WVS survey has a number of questions regarding opinions on international organizations, which can be considered technical knowledge, particularly compared to the LGPI which focuses on local experiences.  WVS-sourced questions are also more likely to return Code 10--Leading/Biasing ($\beta = 0.15, p \le 0.001$). This is likely driven by the structure of the WVS questions which often make a declarative statement such as ``To what extent do you agree with the following statement: The only acceptable religion is my religion." and ask the respondent if they agree or disagree. Since the question starts from a specific point of view, the AI often comments that the question may be leading. Finally, WVS questions are less likely to have Code 4--Unfair Presumption ($\beta = -0.13, p \le 0.01$) compared to the LGPI.  A review of the AI output shows that the Code 4 result is driven by the fact that many of the LGPI questions occur in batteries that use skip patterns. The AI would provide a comment such as ``The phrasing assumes that everyone who is reading the question has not obtained a passport" as it did not know there was an earlier filtering question in the survey.  

\subsubsection{Structure of Output}
A statement order analysis indicates that the AI generally produces statements coded as 5--Double barreled, 1--Vague, and 3--Syntax first, on average placed at 2.39, 2.50, and 2.59 respectively. Codes 7--Difficult recall and 9--Sensitivity are more likely to be placed later, on average, placed at 3.98 and 4.15 respectively. The SysVar code, which indicates statements containing off-task content, were more likely to occur towards the end of the set of 5 statements with an average placement at 4.24. This could suggest that the AI was reaching to fill the 5 slots as requested by the prompt, indicating the need to provide instructions to the AI for cases where no meaningful comments are available. 

\subsubsection{Comparison to Human Performance}
\begin{figure}[htbp]
    \centering
    \includegraphics[width=0.85\textwidth]{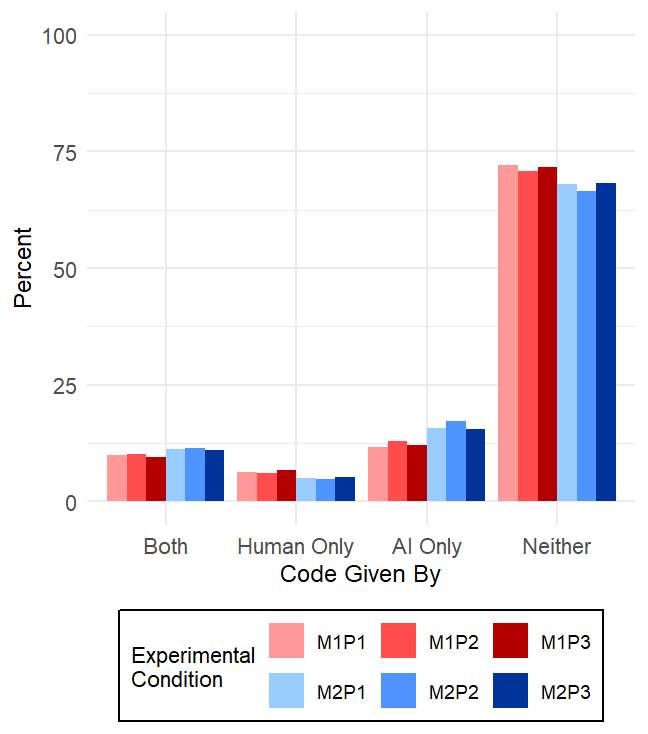}
    \caption{Comparison of Codes between Human Assessment and Experimental Output. In the legend, M1 = GPT--3.4, M2 = GPT--4, P1 = No Persona, P2 = Survey Design Expert , and P3 = Linguist.}
    \label{fig:comparison}
\end{figure}

We consider how the AI-generated codes compare to a human expert,\footnote{The coding was completed by author Erica Ann Metheney.} applying the codebook to the 262 questions. Figure \ref{fig:comparison} shows the variation in how often the human and AI made the same or different determination for each code--question pair. We see that the human and AI output agree--both have the code as present or absent--the majority of the time, around 80\% for all experimental conditions. We find that, for all experimental conditions, rarely did the human code something the AI did not, about 5\% of instances. On the contrary, the AI much more often (around 15\% of instances) coded something the human did not.  

Looking at the differences between M1 (GPT--3.5) and M2 (GPT--4), we see that the percentage of codes missed by the AI (Human Only) decreases by about 1 percentage point and we see a similar increase in the percentage of instances where both gave a code (Both). Although this is a promising finding regarding the improvement of the AI models, we also see that GPT--4 had a higher percentage of ``AI Only" instances where it found things the human did not, with a simultaneous reduction in instances whether neither gave a code (Neither), suggesting possible spurious codes. 

Examining the variation by code, we see a higher percentage of ``AI Only" instances for Code 4--Unfair presumption and 6--Undefined reference period. We also see that there is an increase in ``AI Only" instances by GPT--4 in Code 10--Leading/biasing. The highest percentage of positive agreement (Both)\footnote{Excluding Code 1--Vague Term that was coded almost 100\% of the time by the AI models.} is for Code 4--Unfair presumption. Full results available in Appendix \ref{Appendix:FullResults}. 

\section{Discussion}
Given the plethora of preliminary findings, there is little doubt that AI-based tools and solutions will change how we design, implement, and analyze surveys.

Our study provides consistent evidence that corroborates the intuition that AI can be a powerful tool to support the development of survey questions.  We find that, without any additional model training or parameter adjustments, one can receive feedback on their survey questions that aligns with best practices and recommendations found in the survey methodology literature. Our study also uncovers important sources of variation in the type of feedback received. We find that both choice of model and persona specification influence the content and focus of the feedback. Validity checks show that the AI models generally produce feedback as one would expect, given the nature of the questions provides structured output in a consistent manner and provides feedback similar to that of a human expert. 

While much of our findings support the belief that ``newer models are better", we also find evidence to the contrary. We see that both models can handle issues related to semantics; however, GPT--4.0 was better equipped to identify ``task difficulties," those related to recall, reporting, readability, and recording. The newer model, GPT--4, identified more problems for each question on average, was less likely to produce NOTA, and was more likely to produce nearly all codes, especially issues related to sensitivity and leading. However, our comparison to human output shows that it also tended to flag ``problems" that a human did not find concerning.  

Additionally, like model choice, we find that adequate prompt engineering sharpens the AI content to be more intuitive and relevant. For example, the Linguist persona is more likely to identify question issues related to syntax and the Survey Design Expert is more likely to identify issues such as multi-barreled subjects and answer set problems.  

This study also suggests promising new research avenues for how generative AI can support survey development. The content of NOTA statements shows that generative AI can likely provide a wider range of feedback for question batteries, answer sets, and survey logic, rather than issues related to a single survey question. Outside of the instructions, the AI produced statements related to priming, survey logic, and cognitive load if the survey were given verbally. Furthermore, the AI offered feedback related to the analysis of results but with fielding implications, e.g., potential control variables to account for systematic variation in responses. 

\subsection{Lessons Learned}
Although we find the AI models to be remarkably helpful with survey question refinement, we conclude this paper with lessons we should carry with us as we integrate AI into our survey development workflows and see AI-based survey tools become more common.  Firstly, it is critical that we continue to evaluate AI-based tools and procedures as new models and model versions become available. While we are inclined to believe that newer models will bring better performance, our results indicate that this is not guaranteed to be true across all facets of the survey development process. Similarly, we find strong evidence, consistent with research in other disciplines, that persona and other features of prompt design strongly impact the output content and format, underscoring the need for careful and intentional prompt design. Lastly, we must continually caution ourselves against using AI as a substitution for methodological expertise. We found that the AI models often flagged issues the human coder did not, demonstrating a possibility for new or missed insights, but also potential for spurious observations that take precious time to sort through. 

Our experiment shows that, even today, without any additional training, ChatGPT, and likely other generative AI models, can function as an effective safety net in survey development, providing critical feedback that might otherwise have leave survey data unusable due to a lack of time, financing, or expertise. At the same time, we find that the AI can produce a significant amount of irrelevant feedback and miss issues a human expert would be able to identify. Therefore, for the time being, generative AI is best viewed as an evolving complement to, not a replacement for, established best practices in survey development.  By utilizing these tools within the total survey error framework, we can maximize AI's current strengths, while preparing for more revolutionary applications in the near future. By thinking of AI as a safety net and not a substitution, students, practitioners, and scholars can now apply an additional layer of quality control, catching errors that might otherwise slip past in time- or resource- constrained scenarios.

\clearpage
\appendix 
\appendixpage

\begin{appendices}

\section{Sourcing Experimental Units}\label{Appendix:SourcingEUs}
We sourced 262 questions from surveys developed by Gallup, World Values Survey Association (WVS), and the Governance and Local Development Institute (GLD). Specific questions from Gallup, WVS, and LGPI were selected on grounds of their ability to be posed as ``stand alone" questions. Questions were excluded if they relied on answer set structure, survey logic, were not phrased as a question, or were redundant. Questions that were structured as fill in the blank (for example, ''I am...with my government.") were also excluded. 

These sources were partially selected to explore if the AI can handle questions with a range of expected quality. From Gallup, we selected the publicly available Q12 survey containing 13 questions to measure employee engagement \parencite{gallup_overview}. We selected questions from the 7th wave of WVS. This survey has been translated and implemented in over 90 countries since the first wave in 1981 \parencite{wvs_methodology}. Finally, the 2019 version of the LGPI survey conducted by GLD was implemented in 3 countries requiring six translations \parencite{gld_overview}. We believe these three sources provide a range of refinement and topics. Of our externally sourced questions, we expect the AI to identify the most potential  issues for LGPI because this survey covers the largest range of topics with the fewest iterations. GLD is a newer organization than Gallup and WVS with lower capacity and experience to resolve translation issue within their questions. Thus, we anticipate fewer problems to be flagged by the AI for the WVS questions because WVS had six prior waves of experience with international surveys, translated into more languages than the LGPI, and covers fewer topics. Finally, the short precise Q12 survey from Gallup is the mostly widely used and vetted compared to WVS and LGPI and thus should have the fewest unresolved problems that the AI should be able to identify.

\section{Qualitative Coding of GPT Output}\label{Appendix:Qual}

Qualitative coding was done in two different stages. In the first round, the output from the questionable questions were coded during codebook refinement, and the authors knew the source though not the treatment or question. We used an abductive approach which sits at the intersection of inductive and deductive approaches to analyze the output from the AI \parencite{alvesson2022reflexive}. We begin with a theory driven framework to see if the AI can identify what the literature highlights as important while leaving room for emergent patterns. For the first 100 statements, the authors coded together to decide a code and settle on the proper line of reasoning. Then in two batches of 100, the authors coded independently and settled discrepancies while updating the code book. With the final 300 questionable question statements, the authors coded independently and passed an inter-coder reliability check \footnote{The coders were above 80\% agreement for all codes and 70\% full agreement on statements.}. 

For the second round, the output from the other sources were randomized and anonymized at the statement level in an attempt to blind the qualitative coders from the source, treatment, and question. The authors coded the first 100 together to ensure the codebook's applicability for output from a different question sources. Approximately a thousand statements were coded by two trained qualitative coders, though due to low inter coder reliability all codes were double checked by the authors. Once this was clarified, authors coded the remainder of the output independently and highlighted and decided codes for material that was unclear or novel. 

\begin{landscape}

\begin{longtable}[t]{|p{4em}|p{3em}|p{6em}|p{14em}|p{8em}|p{12em}|} 
\caption{Complete Qualitative Codebook}
\label{tab:codescheme2}\\
\toprule
 \textbf{Problem Type}&\textbf{Code} & \textbf{Label} & \textbf{Description} & \textbf{Problem might be solved by:}& \textbf{Example(s):} \\ \hline 
 Question Content&1& Vague term or phrase & When a phrase or term is unclear, respondents don’t know what you mean. Respondents don’t know which aspect of the term to answer about, respondents don’t know what you’re talking about & Definition or quantifiers, Telling   participants where to look or think  & Kids, car, own, etc.  \\ \hline 
 &2             & Specialized knowledge                                      & When a term or phrase requires more specialized information, terminology. Quality of response is affected by respondent’s knowledge of the topic                                                                                                                         & Additional uniform, technical information, or the respondent being an expert already OR Filtering out people without the   specialized knowledge & PCP as angeldust or TIA as microstroke                                                                                                                                                                                                                                                                                                                                                                                                                             \\ \hline 
 Question Structure&3             & Syntax problem                                             & Structural problems like question too long, complex, or awkward, grammar problems                                                                                                                                                                                        & Rewrite the question                                                                                                                             & If any at all, verbs not matching the   time frame                                                                                                                                                                                                                                                                                                                                                                                                                 \\ \hline 
 &4             & Unfair Presumption                                         & The questions assumes to a detriment prior   knowledge or conditions about the respondent                                                                                                                                                                                & Filter or acknowledge exclusion                                                                                                                  & Assumes respondent knowledge, ability to get pregnant                                                                                                                                                                                                                                                                                                                                                                                                              \\ \hline 
 &5             & Double Barrelled                                           & Several questions or multiple subjects                                                                                                                                                                                                                                   & One subject or multiple questions                                                                                                                & Women and children, courts and police                                                                                                                                                                                                                                                                                                                                                                                                                              \\ \hline 
 Affiliated with time&6             & Undefined or ill-defined reference period                  & Question doesn’t sufficiently define or narrow the time period the respondent should consider                                                                                                                                                                            & Providing specific time frame                                                                                                                    & Recent, week, prior                                                                                                                                                                                                                                                                                                                                                                                                                                                \\ \hline 
 &7             & Difficult recall or reference period                       & Time frame accurately specified but is too specific or far in the past to be easy to answer. this is also highly specific or confusing reference period. Difficult to recall the specific item or  topic of interest in the question (eg. too many options to pick from) & Provide memory cues (Simple orienting memory info)                                                                                               & Election of 2001 (Bush v. Al Gore election); What did you have for breakfast 10 days ago v. Where were you for 9/11?                                                                                                                                                                                                                                                                                                                                               \\ \hline 
 Judgment/ Evaluation&8             & Complex estimation/computation                             & Questions with sizable calculations (math) or combination of lots of information                                                                                                                                                                                         & Defining criteria or orienting to values                                                                                                         & Grams of butter over a week, perceived   speed of rate increase                                                                                                                                                                                                                                                                                                                                                                                                    \\ \hline 
 &9             & Potentially Sensitive/Emotionally charged or controversial & Question makes people uncomfortable, taboo, or offended, triggers emotion                                                                                                                                                                                                & Not asking the question                                                                                                                          & Obamacare, formality issue                                                                                                                                                                                                                                                                                                                                                                                                                                         \\ \hline 
 &10            & Leading/Biasing question                                   & The wording of the question causes people to answer differently than their true opinion. The wording of the question makes certain answer choices or types of answer more likely than they would be if the   question was worded differently or used different words.    & Rephrasing the question; Randomizing part of the question or randomizing the answer set (if also answer set issue)                               & Excellent listed first making people more likely to choose excellent.  List of answer is so long, people don’t read all options and choose items earlier in the list. Multiple topics are presented in the question and respondents might lock on to the first topic, ignoring the rest.  Question is about a controversial topic and people will feel compelled to answer a certain way so they are not judged,   e.g. Would you like to help starving children?  \\ \hline 
 Emergent codes&11            & Answer set                                                 & If the question had some answers within it and the AI flags problems affiliated                                                                                                                                                                                          & Modifying answer set                                                                                                                             & No neutral option, order of answers, insufficient answers                                                                                                                                                                                                                                                                                                                                                                                                          \\ \hline 
               && Internal Code: Systematic Variation                        & Specific types/demographics of people will consistently answer certain ways                                                                                                                                                                                              & Not a problem for question but of interest for analysis                                                                                          &                                                                                                                                                                                                                                                                                                                                                                                                                                                                    \\ \hline 
               && None of the above                                          & Problems that AI flags inappropriately   (like cultural or linguistic variation) OR problems affiliated with the survey’s   logic, format, mode, or context, this includes if you need a “treatment   paragraph”                                                         & Not a problem for questions, solved by   survey or “treatment paragraph”                                                                         & Cultures are different and may answer   differently, it’s difficult to ask questions on the phone, context of why   you’re asking about teachers or why you should care      \\ \bottomrule
   \end{longtable}
\end{landscape}

\begin{table}[htbp]
\centering
\caption{Rothgeb et al (2007) adapted coding scheme. Asterisk indicates control question for experiment. Codes marked with ``+" indicates emergent codes identified by project researchers.}\label{tab:codescheme1}
\begin{tabular}{|p{6em}p{1em}p{10em}|p{17em}|}
\hline
\textbf{Type} &    & \textbf{Description} & \textbf{Example} \\ \hline
Question Content      & 1  & Vague/Undefined Topic/Term  & Do you own a car?* \\ \cline{2-4} 
                     & 2  & Specialized Knowledge (solvable by additional information) & Do you think that adults should be able to use PCP without any legal penalty?*  \\  \hline
Question Structure   & 3  & Question too long, Complex or Awkward, Syntax Problem                           & Before you got married, how long did you live in Maryland after you graduated from college?*                                                                                                                             \\ \cline{2-4} 
                     & 4  & Unfair Presumption                               & What is the best thing about driving a convertible?                                                                                                                                                                                                                       \\ \cline{2-4} 
                     & 5  & Several Questions or Multiple Subjects              & Do you think women and children should be given the first available flu shots?*                                                                                                                                         \\ \cline{2-4} 
                     & 6  & Leading/Biasing Question+                       &                                                                                                                                                                               \\ \cline{2-4} 
                     & 7  & Unclear/Vague Question Scope+                       &                                                                                                                                                                                \\ \cline{2-4} 
                     & 8  & Undefined Reference Period+                         & How long have you lived in College Park?*                                                                                                                                                                               \\ \hline
                     
Retrieval from Memory & 9  & Difficult Recall/Reference Period or Lack of Memory Cues                            & Which candidate did you vote for in the presidential election of 2004?                                                            \\ \hline
Judgment/ Evaluation                     & 10 & Complex Estimation                                 & Compared to a year ago, do you feel the prices of most things you buy are going up faster than they did then, going up as fast, going up slower, or not going up at all?*                                                \\ \cline{2-4} 
                     & 11 & Potentially sensitive, Emotionally Charged, or Controversial                     &                                                                                                                                                                                                                        
                                                                                            \\ \hline
Internal Codes                    & I1 &  Answer Set Problem+   &                                                \\ \cline{2-4}
                & I2 &  Systematic Variation in Responses+   &                                                \\ \hline
    
\end{tabular}
\end{table}

\textbf{Typical combinations:}
\begin{itemize}
    \item Assumption of prior (specialized) knowledge: 2 and 4 b/c unfair presumption
    \item Especially problematic words like value are often 1 and 11 because it’s a scope problem money v. emotions, but even if provided a scope still vague for monetary value
    \item 9 and 10 often combined when the emotional trigger leads people to ONE direction in the answer (for example triggering pride)
\end{itemize}

\textit{Additional notes}
\begin{itemize}
    \item For most questions, we do not include any answer sets. This means that if it says “no answer set provided or unclear how respondents should answer” this would be “none of the above” because it’s beyond the scope of the project. But for some questions, we do have some answers like “agree or disagree with the following…” and if this is the case and the AI mentions it, it’s most likely 12 (meaning that you can assume if it looks like an answer set it is 12). It may also be other things for example (The questions structure, starting with the positive term excellent, might subtly encourage respondents to think more positively about the performance of the police and courts. By starting with the most positive option, it may create a priming effect, leading respondents to consider the positive aspects first.)
    \item Cultural differences or variation are typically none of the above, but read closely if it mentions normative, moral, religious, personal issues because that’s within the scope and is coded 9 for sensitivity concerns.
    \item When unsure, think about how to fix the problem.
    \item If multiple codes and unsure, try to break it down by sentence.
    \item Air on the side of more codes, but if you code more than 1 you have to find explicit proof in the output.
\end{itemize}

\textbf{Changes during codebook refinement}
\begin{itemize}
\item Deleted all codes related to response selections, interview difficulties, and survey flow because it’s beyond the scope of this study.
\item Changed the terminology from ``complex” to technical. 
\item Combine vague topic and undefined/vague term because they are very similar and the ``solution” for the researcher is the same, i.e. to specify the topic. 
\item Combine question too long and complex or awkward syntax because they are very similar and the “solution” for the researcher is the same, i.e. rewrite the question. 
\item Narrowed/redefined “erroneous” to mean more of a presupposed condition. (ex. What’s the best thing about driving a convertible? Presupposes the subject has access to a convertible and enjoys driving it.) This problem might not exist if given the whole survey flow, but is especially relevant given the individual, independent question. 
\item ``Long” recall/reference period is changed to difficult because there is something different about ``what day was your child born” and ``what did you have 2 weeks ago today for breakfast?”
\end{itemize}

\section{Full Results}\label{Appendix:FullResults}

\subsection{Power Analysis}

\begin{figure}[htbp]
\centering
\includegraphics[width=\textwidth]{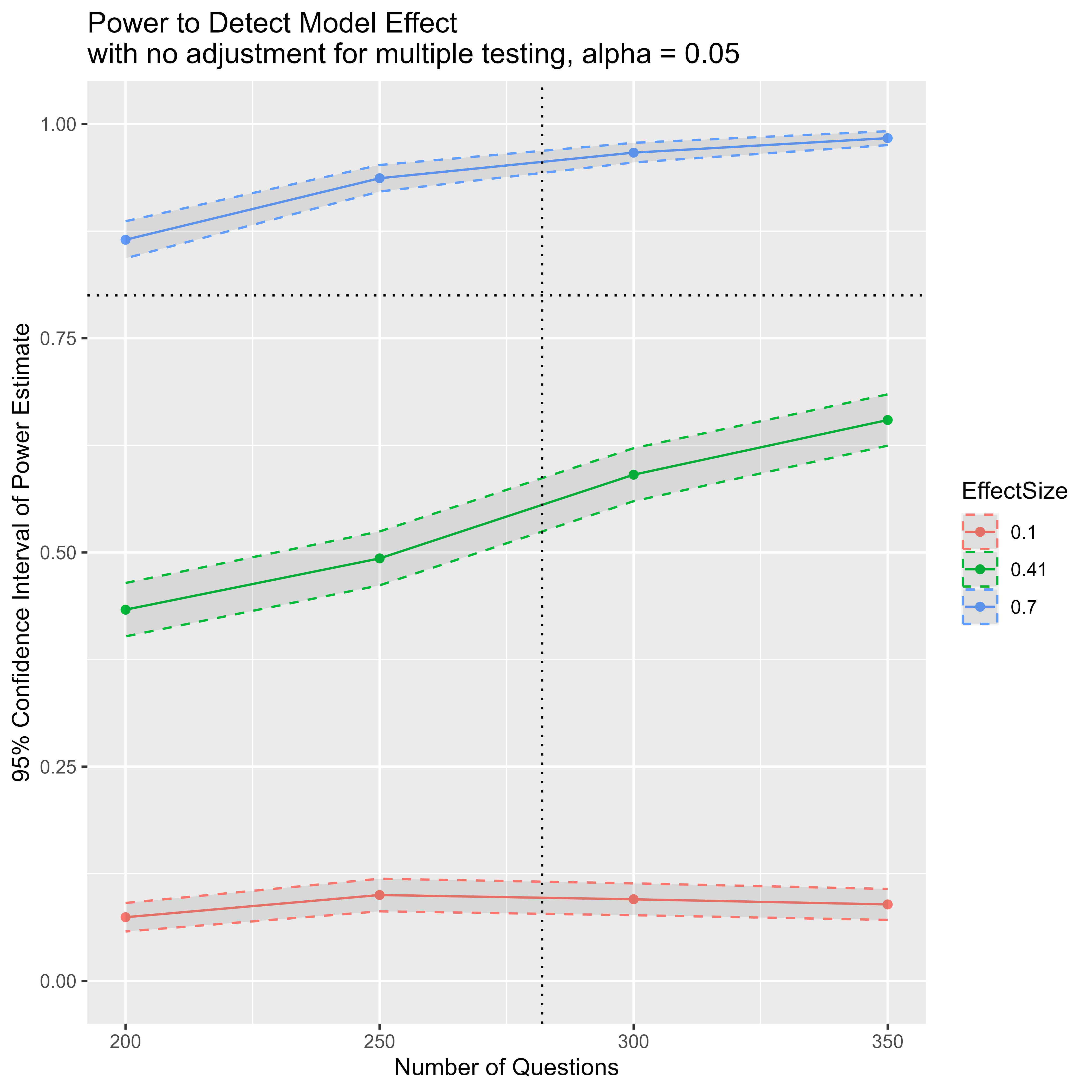}
\caption{Power analysis results for detecting a Model effect of sizes 0.1, 0.41, and 0.7 (without multiple testing correction)}
\end{figure}

\begin{figure}[htbp]
\centering
\includegraphics[width=\textwidth]{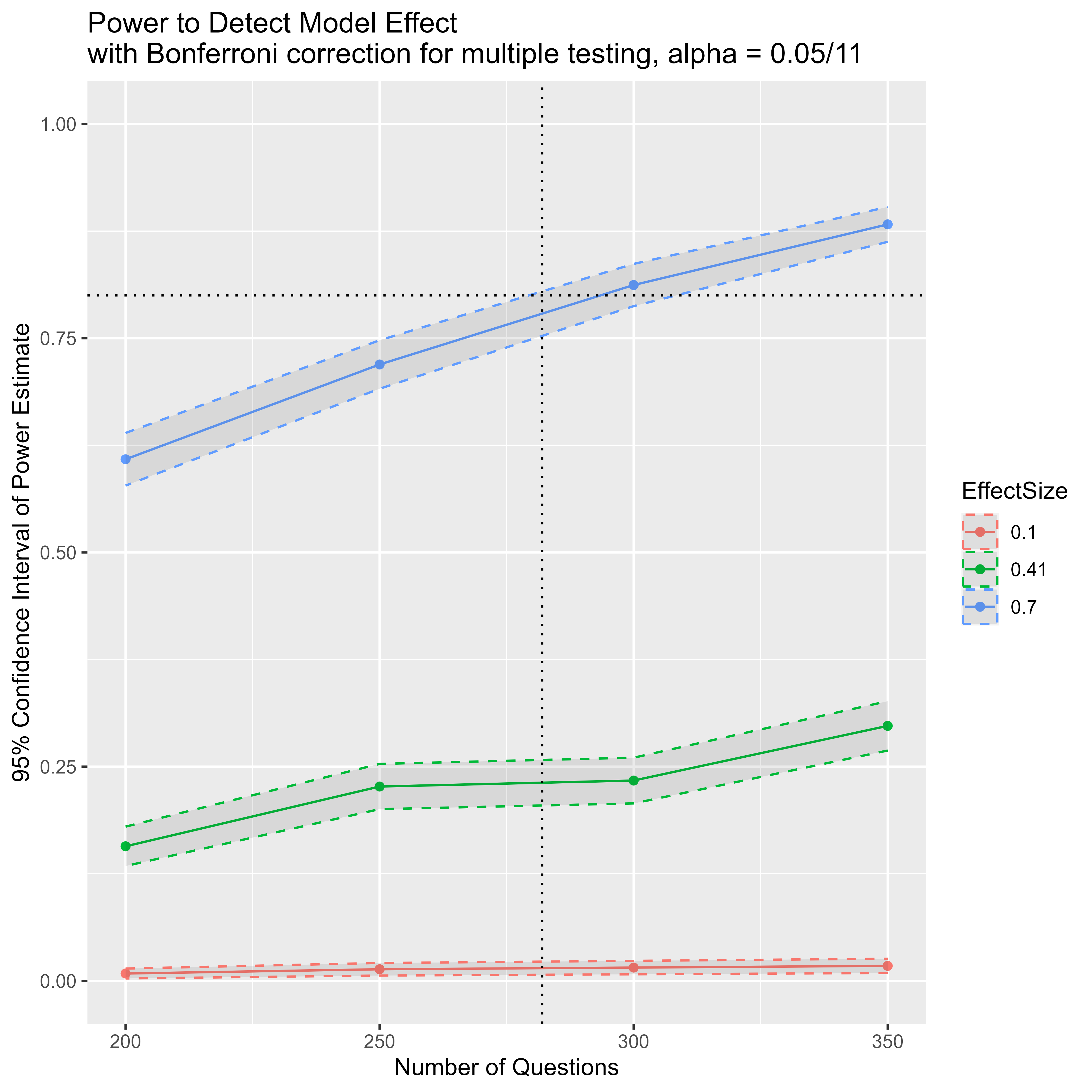}
\caption{Power analysis results for detecting a Model effect of sizes 0.1, 0.41, and 0.7 (with multiple testing correction)}
\end{figure}

\begin{figure}[htbp]
\centering
\includegraphics[width=\textwidth]{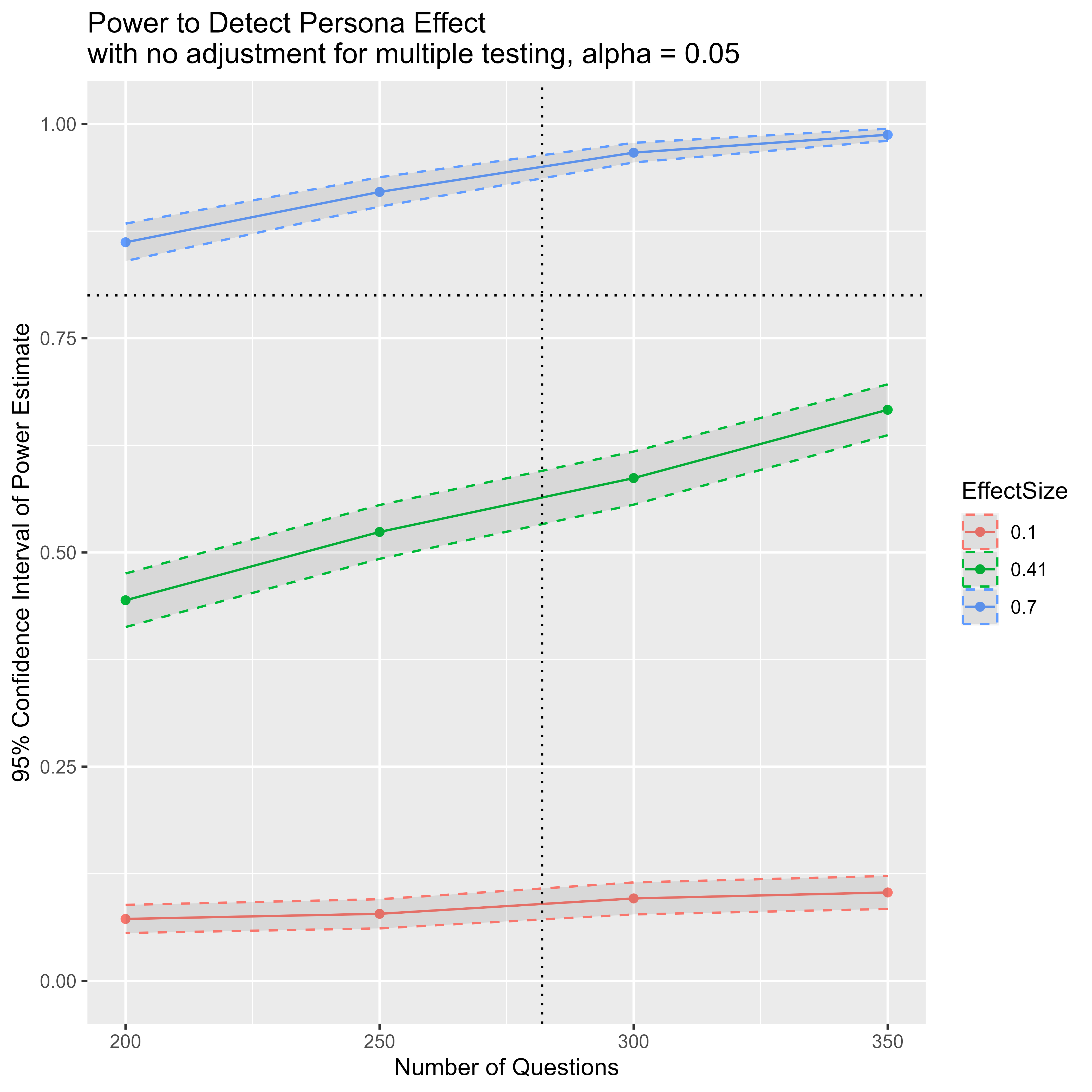}
\caption{Power analysis results for detecting a Persona effect of sizes 0.1, 0.41, and 0.7 (without multiple testing correction)}
\end{figure}

\begin{figure}
\centering
\includegraphics[width=\textwidth]{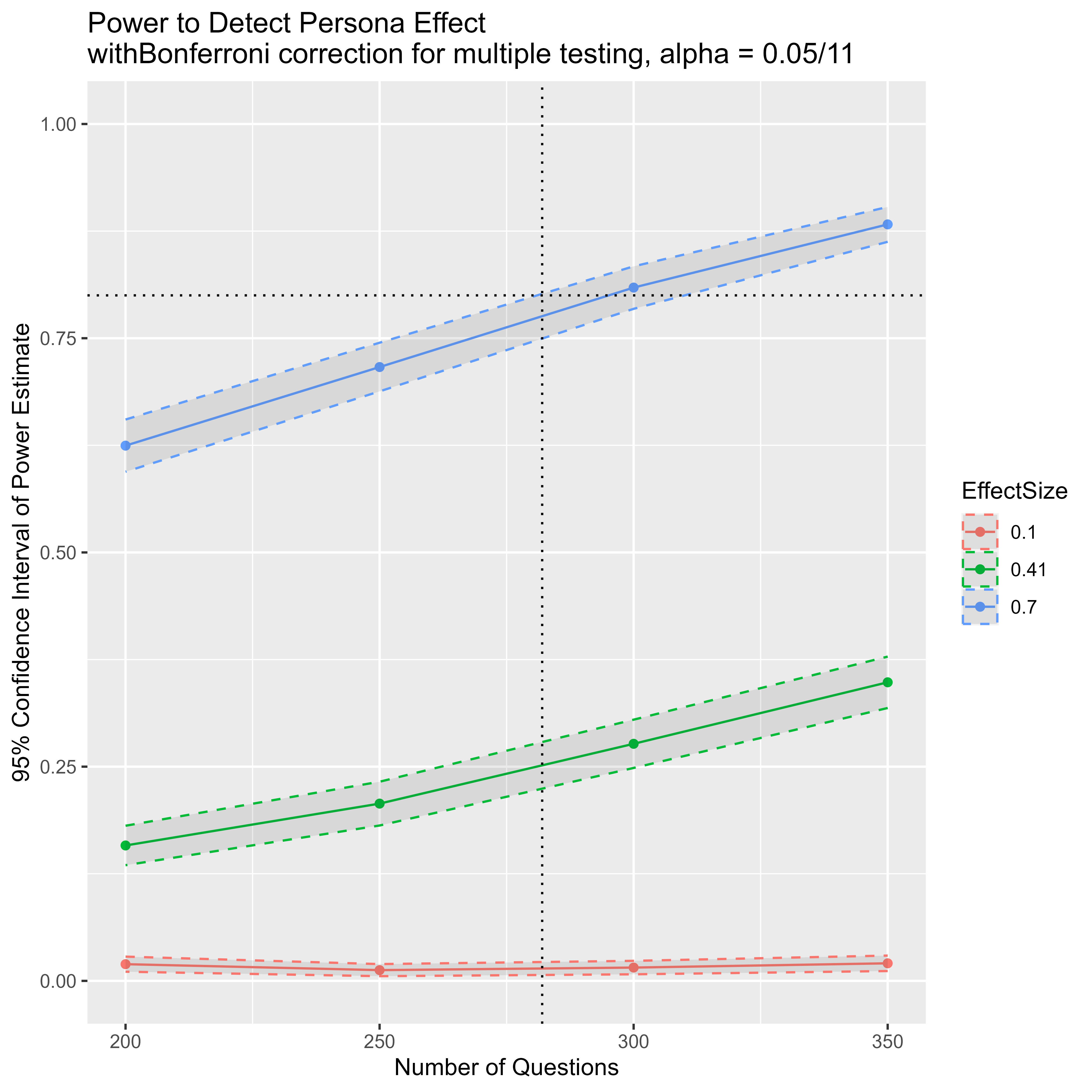}
\caption{Power analysis results for detecting a Persona effect of sizes 0.1, 0.41, and 0.7 (with multiple testing correction)}
\end{figure}

\begin{figure}[htbp]
\centering
\includegraphics[width=\textwidth]{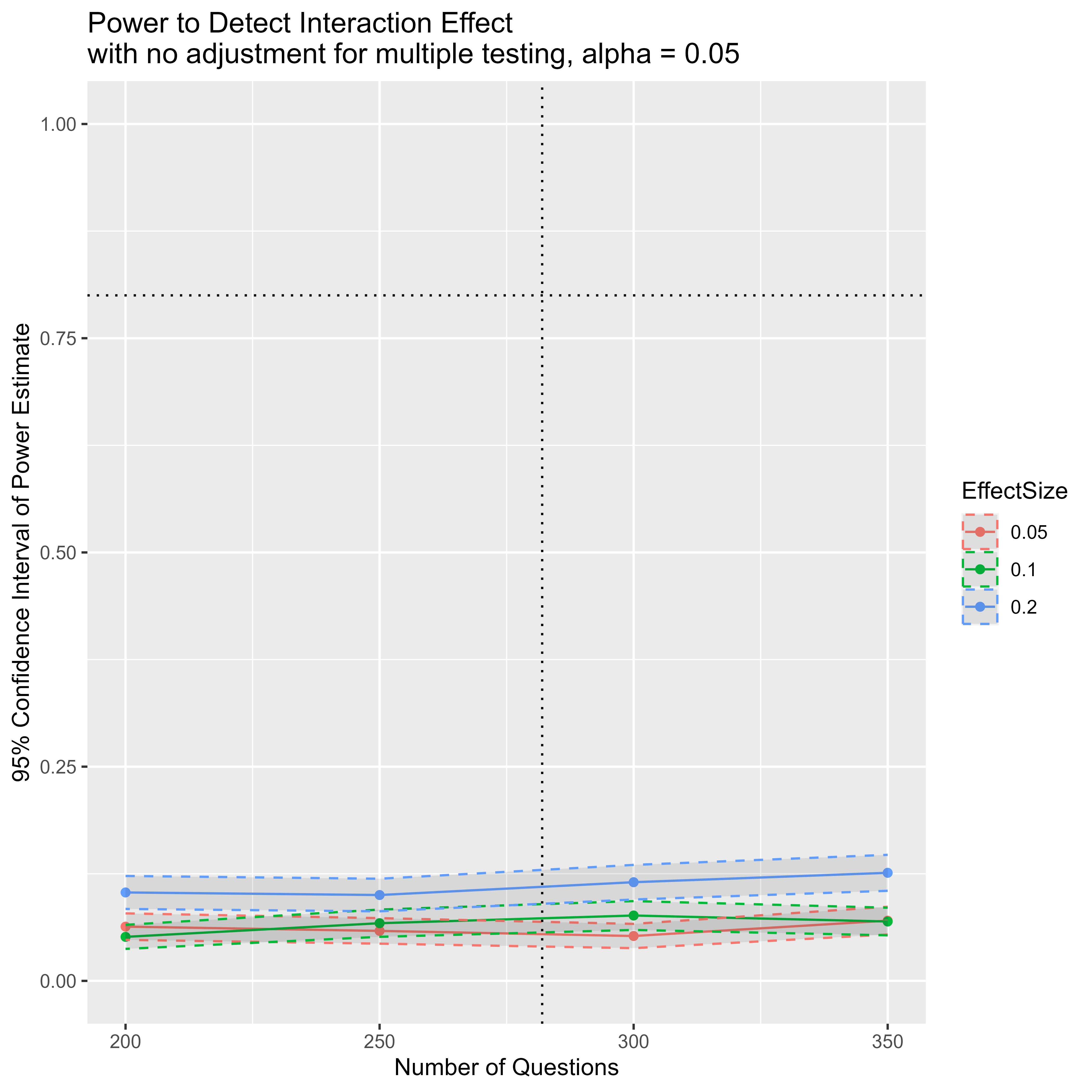}
\caption{Power analysis results for detecting an interaction effect between Model and Persona of sizes 0.05, 0.1, and 0.2 (without multiple testing correction)}
\end{figure}

\begin{figure}[htbp]
\centering
\includegraphics[width=\textwidth]{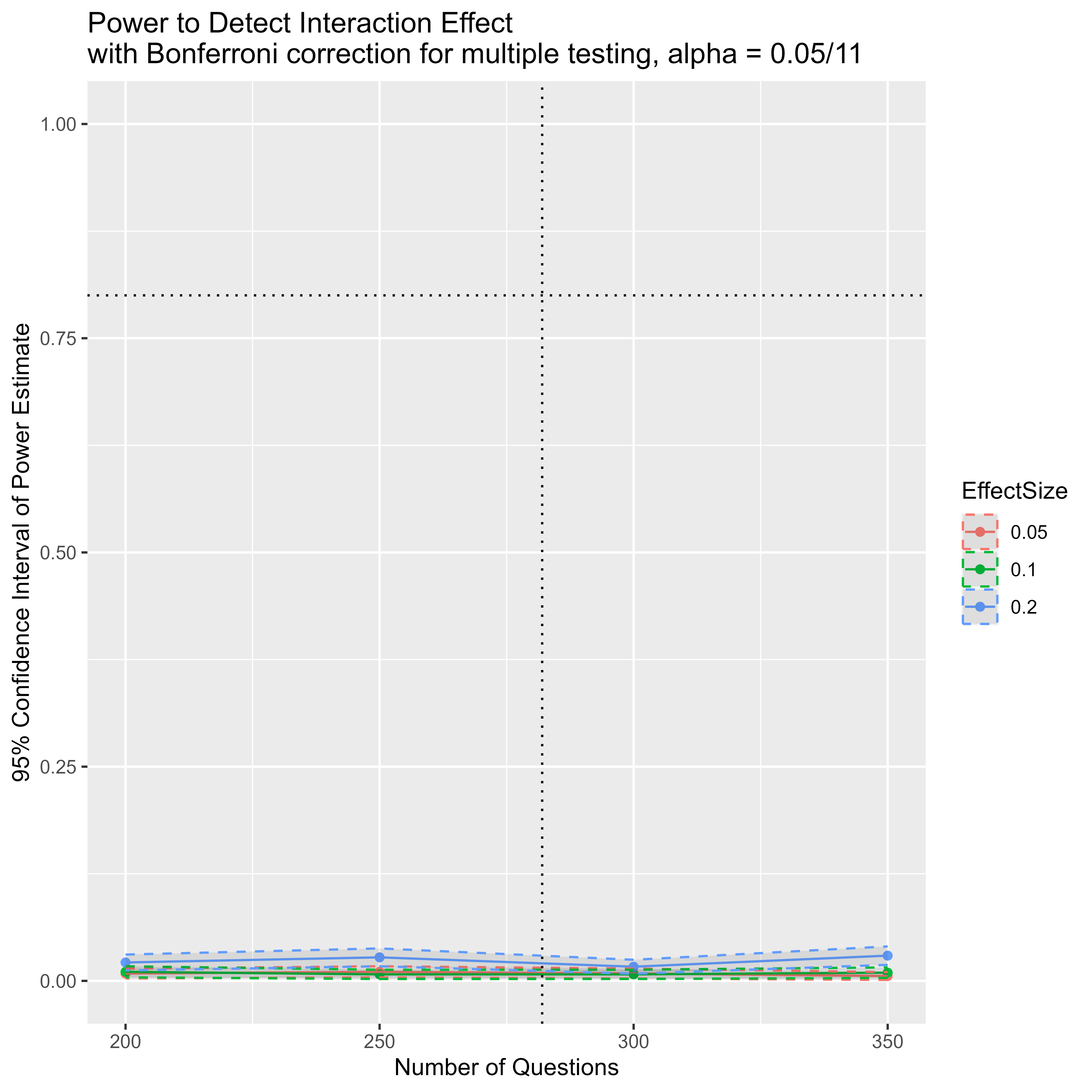}
\caption{Power analysis results for detecting an interaction effect between Model and Persona of sizes 0.05, 0.1, and 0.2 (with multiple testing correction)}
\end{figure}

\subsection{Experimental Analysis}

\begin{table}[htbp]
\caption{Multilevel Regression Results for Number of Codes Per Treatment-Question Pair}
\begin{center}
\begin{tabular}{l c}
\hline
 & Number of Codes \\
\hline
(Intercept)               & $2.34^{***}$ \\
                          & $(0.05)$     \\
ModelM2                   & $0.55^{***}$ \\
                          & $(0.04)$     \\
PersonaP2                 & $0.22^{***}$ \\
                          & $(0.05)$     \\
PersonaP3                 & $-0.03$      \\
                          & $(0.05)$     \\
\hline
AIC                       & $4117.60$    \\
BIC                       & $4149.76$    \\
Log Likelihood            & $-2052.80$   \\
Num. obs.                 & $1571$       \\
Num. groups: Qs     & $262$        \\
Var: Question (Int.) & $0.18$       \\
Var: Residual             & $0.67$       \\
\hline
\multicolumn{2}{l}{\scriptsize{$^{***}p<0.001$; $^{**}p<0.01$; $^{*}p<0.05$}}
\end{tabular}
\label{tab:NumCodesTrtQ}
\end{center}
\end{table}

\begin{landscape}
    \begin{table}[htbp]
\caption{Regression Results with Holmes Multiple Testing Correction on Model Effect Only}
\label{tab:regM2corrected}
\begin{center}
\begin{tabular}{l c c c c c c c c c c c}
\hline
 & Code 1 & Code 2 & Code 3 & Code 4 & Code 5 & Code 6 & Code 7 & Code 8 & Code 9 & Code 10 & Code 11 \\
\hline
(Intercept) & $0.98^{***}$ & $0.03^{**}$ & $0.05^{**}$  & $0.45^{***}$ & $0.09^{***}$ & $0.40^{***}$ & $0.03^{**}$ & $0.00$     & $0.07^{***}$ & $0.06^{***}$ & $0.18^{***}$ \\
            & $(0.01)$     & $(0.01)$    & $(0.02)$     & $(0.03)$     & $(0.02)$     & $(0.03)$     & $(0.01)$    & $(0.00)$   & $(0.02)$     & $(0.02)$     & $(0.02)$     \\
ModelM2     & $0.01$       & $0.01$      & $0.08^{***}$ & $0.02$       & $0.06^{***}$ & $0.04$       & $0.02$      & $0.01^{*}$ & $0.14^{***}$ & $0.15^{***}$ & $0.03$       \\
            & $(0.00)$     & $(0.01)$    & $(0.01)$     & $(0.02)$     & $(0.01)$     & $(0.02)$     & $(0.01)$    & $(0.00)$   & $(0.01)$     & $(0.02)$     & $(0.02)$     \\
PersonaP2   & $-0.00$      & $0.02^{*}$  & $0.03$       & $0.03$       & $0.05^{**}$  & $0.02$       & $0.00$      & $-0.00$    & $-0.02$      & $0.04$       & $0.06^{**}$  \\
            & $(0.01)$     & $(0.01)$    & $(0.02)$     & $(0.03)$     & $(0.02)$     & $(0.02)$     & $(0.01)$    & $(0.00)$   & $(0.02)$     & $(0.02)$     & $(0.02)$     \\
PersonaP3   & $0.00$       & $0.01$      & $0.05^{**}$  & $-0.01$      & $-0.01$      & $-0.04$      & $-0.00$     & $0.00$     & $0.00$       & $-0.01$      & $-0.02$      \\
            & $(0.01)$     & $(0.01)$    & $(0.02)$     & $(0.03)$     & $(0.02)$     & $(0.02)$     & $(0.01)$    & $(0.00)$   & $(0.02)$     & $(0.02)$     & $(0.02)$     \\
\hline
AIC                       & -2662.01   & -752.65   & 862.20     & 2136.58    & 867.48     & 1786.64    & -869.76   & -3803.92  & 895.13     & 1017.60    & 1425.46       \\
BIC                       & -2629.86   & -720.49   & 894.36     & 2168.74    & 899.64     & 1818.79    & -837.61   & -3771.77  & 927.28     & 1049.76    & 1457.62       \\
Log Likelihood            & 1337.01    & 382.32    & -425.10    & -1062.29   & -427.74    & -887.32    & 440.88    & 1907.96   & -441.56    & -502.80    & -706.73     \\
Num. obs.                 & 1571       & 1571      & 1571       & 1571       & 1571       & 1571       & 1571      & 1571      & 1571       & 1571       & 1571            \\
Num. groups: Qs     & 262        & 262       & 262        & 262        & 262        & 262        & 262       & 262       & 262        & 262        & 262              \\
Var: Question (Int.) & 0.00       & 0.02      & 0.02       & 0.06       & 0.03       & 0.11       & 0.01      & 0.00      & 0.02       & 0.02       & 0.05         \\
Var: Residual             & 0.01       & 0.03      & 0.09       & 0.19       & 0.08       & 0.13       & 0.03      & 0.01      & 0.09       & 0.09       & 0.11            \\
\hline
\multicolumn{12}{l}{\scriptsize{$^{***}p<0.001$; $^{**}p<0.01$; $^{*}p<0.05$}}
\end{tabular}
\end{center}
\end{table}
\end{landscape}

\begin{landscape}
\begin{table}[htbp]
\caption{Regression Results with Holmes Multiple Testing Correction on Persona Effect (P2 = Survey Design Expert) Only}
\label{tab:regP2corrected}
\begin{center}
\begin{tabular}{l c c c c c c c c c c c}
\hline
 & Code 1 & Code 2 & Code 3 & Code 4 & Code 5 & Code 6 & Code 7 & Code 8 & Code 9 & Code 10 & Code 11 \\
\hline
(Intercept) & $0.98^{***}$ & $0.03^{**}$ & $0.05^{**}$  & $0.45^{***}$ & $0.09^{***}$ & $0.40^{***}$ & $0.03^{**}$ & $0.00$      & $0.07^{***}$ & $0.06^{***}$ & $0.18^{***}$ \\
            & $(0.01)$     & $(0.01)$    & $(0.02)$     & $(0.03)$     & $(0.02)$     & $(0.03)$     & $(0.01)$    & $(0.00)$    & $(0.02)$     & $(0.02)$     & $(0.02)$     \\
ModelM2     & $0.01$       & $0.01$      & $0.08^{***}$ & $0.02$       & $0.06^{***}$ & $0.04^{*}$   & $0.02^{*}$  & $0.01^{**}$ & $0.14^{***}$ & $0.15^{***}$ & $0.03$       \\
            & $(0.00)$     & $(0.01)$    & $(0.01)$     & $(0.02)$     & $(0.01)$     & $(0.02)$     & $(0.01)$    & $(0.00)$    & $(0.01)$     & $(0.02)$     & $(0.02)$     \\
PersonaP2   & $-0.00$      & $0.02$      & $0.03$       & $0.03$       & $0.05^{*}$   & $0.02$       & $0.00$      & $-0.00$     & $-0.02$      & $0.04$       & $0.06^{*}$   \\
            & $(0.01)$     & $(0.01)$    & $(0.02)$     & $(0.03)$     & $(0.02)$     & $(0.02)$     & $(0.01)$    & $(0.00)$    & $(0.02)$     & $(0.02)$     & $(0.02)$     \\
PersonaP3   & $0.00$       & $0.01$      & $0.05^{**}$  & $-0.01$      & $-0.01$      & $-0.04$      & $-0.00$     & $0.00$      & $0.00$       & $-0.01$      & $-0.02$      \\
            & $(0.01)$     & $(0.01)$    & $(0.02)$     & $(0.03)$     & $(0.02)$     & $(0.02)$     & $(0.01)$    & $(0.00)$    & $(0.02)$     & $(0.02)$     & $(0.02)$     \\
\hline
AIC                       & -2662.01   & -752.65   & 862.20     & 2136.58    & 867.48     & 1786.64    & -869.76   & -3803.92  & 895.13     & 1017.60    & 1425.46       \\
BIC                       & -2629.86   & -720.49   & 894.36     & 2168.74    & 899.64     & 1818.79    & -837.61   & -3771.77  & 927.28     & 1049.76    & 1457.62       \\
Log Likelihood            & 1337.01    & 382.32    & -425.10    & -1062.29   & -427.74    & -887.32    & 440.88    & 1907.96   & -441.56    & -502.80    & -706.73     \\
Num. obs.                 & 1571       & 1571      & 1571       & 1571       & 1571       & 1571       & 1571      & 1571      & 1571       & 1571       & 1571            \\
Num. groups: Qs     & 262        & 262       & 262        & 262        & 262        & 262        & 262       & 262       & 262        & 262        & 262              \\
Var: Question (Int.) & 0.00       & 0.02      & 0.02       & 0.06       & 0.03       & 0.11       & 0.01      & 0.00      & 0.02       & 0.02       & 0.05         \\
Var: Residual             & 0.01       & 0.03      & 0.09       & 0.19       & 0.08       & 0.13       & 0.03      & 0.01      & 0.09       & 0.09       & 0.11            \\
\hline
\multicolumn{12}{l}{\scriptsize{$^{***}p<0.001$; $^{**}p<0.01$; $^{*}p<0.05$}}
\end{tabular}
\end{center}
\end{table}
\end{landscape}

\begin{landscape}
\begin{table}[htbp]
\caption{Regression Results with Holmes Multiple Testing Correction on Persona Effect (P3 = Linguist) Only}
\label{tab:regP3corrected}
\begin{center}
\begin{tabular}{l c c c c c c c c c c c}
\hline
 & Code 1 & Code 2 & Code 3 & Code 4 & Code 5 & Code 6 & Code 7 & Code 8 & Code 9 & Code 10 & Code 11 \\
\hline
(Intercept) & $0.98^{***}$ & $0.03^{**}$ & $0.05^{**}$  & $0.45^{***}$ & $0.09^{***}$ & $0.40^{***}$ & $0.03^{**}$ & $0.00$      & $0.07^{***}$ & $0.06^{***}$ & $0.18^{***}$ \\
            & $(0.01)$     & $(0.01)$    & $(0.02)$     & $(0.03)$     & $(0.02)$     & $(0.03)$     & $(0.01)$    & $(0.00)$    & $(0.02)$     & $(0.02)$     & $(0.02)$     \\
ModelM2     & $0.01$       & $0.01$      & $0.08^{***}$ & $0.02$       & $0.06^{***}$ & $0.04^{*}$   & $0.02^{*}$  & $0.01^{**}$ & $0.14^{***}$ & $0.15^{***}$ & $0.03$       \\
            & $(0.00)$     & $(0.01)$    & $(0.01)$     & $(0.02)$     & $(0.01)$     & $(0.02)$     & $(0.01)$    & $(0.00)$    & $(0.01)$     & $(0.02)$     & $(0.02)$     \\
PersonaP2   & $-0.00$      & $0.02^{*}$  & $0.03$       & $0.03$       & $0.05^{**}$  & $0.02$       & $0.00$      & $-0.00$     & $-0.02$      & $0.04$       & $0.06^{**}$  \\
            & $(0.01)$     & $(0.01)$    & $(0.02)$     & $(0.03)$     & $(0.02)$     & $(0.02)$     & $(0.01)$    & $(0.00)$    & $(0.02)$     & $(0.02)$     & $(0.02)$     \\
PersonaP3   & $0.00$       & $0.01$      & $0.05^{*}$   & $-0.01$      & $-0.01$      & $-0.04$      & $-0.00$     & $0.00$      & $0.00$       & $-0.01$      & $-0.02$      \\
            & $(0.01)$     & $(0.01)$    & $(0.02)$     & $(0.03)$     & $(0.02)$     & $(0.02)$     & $(0.01)$    & $(0.00)$    & $(0.02)$     & $(0.02)$     & $(0.02)$     \\
\hline
AIC                       & -2662.01   & -752.65   & 862.20     & 2136.58    & 867.48     & 1786.64    & -869.76   & -3803.92  & 895.13     & 1017.60    & 1425.46       \\
BIC                       & -2629.86   & -720.49   & 894.36     & 2168.74    & 899.64     & 1818.79    & -837.61   & -3771.77  & 927.28     & 1049.76    & 1457.62       \\
Log Likelihood            & 1337.01    & 382.32    & -425.10    & -1062.29   & -427.74    & -887.32    & 440.88    & 1907.96   & -441.56    & -502.80    & -706.73     \\
Num. obs.                 & 1571       & 1571      & 1571       & 1571       & 1571       & 1571       & 1571      & 1571      & 1571       & 1571       & 1571            \\
Num. groups: Qs     & 262        & 262       & 262        & 262        & 262        & 262        & 262       & 262       & 262        & 262        & 262              \\
Var: Question (Int.) & 0.00       & 0.02      & 0.02       & 0.06       & 0.03       & 0.11       & 0.01      & 0.00      & 0.02       & 0.02       & 0.05         \\
Var: Residual             & 0.01       & 0.03      & 0.09       & 0.19       & 0.08       & 0.13       & 0.03      & 0.01      & 0.09       & 0.09       & 0.11            \\
\hline
\multicolumn{12}{l}{\scriptsize{$^{***}p<0.001$; $^{**}p<0.01$; $^{*}p<0.05$}}
\end{tabular}

\end{center}
\end{table}
\end{landscape}

\subsection{Robustness Checks}
This section shows the results when a logistic multilevel model is used to model the likelihood of each code. 

\begin{landscape}
    \begin{table}
\caption{Logistic Regression Results with Holmes Multiple Testing Correction on Model Effect Only}
\begin{center}
\begin{tabular}{l c c c c c c c c c c c}
\hline
 & Code 1 & Code 2 & Code 3 & Code 4 & Code 5 & Code 6 & Code 7 & Code 8 & Code 9 & Code 10 & Code 11 \\
\hline
(Intercept) & $9.23^{***}$ & $-9.29^{***}$ & $-3.57^{***}$ & $-0.29^{*}$ & $-3.60^{***}$ & $-0.71^{***}$ & $-8.06^{***}$ & $-31.99$  & $-3.61^{***}$ & $-3.60^{***}$ & $-2.30^{***}$ \\
            & $(1.27)$     & $(0.90)$      & $(0.27)$      & $(0.14)$    & $(0.31)$      & $(0.19)$      & $(0.92)$      & $(76.01)$ & $(0.28)$      & $(0.27)$      & $(0.22)$      \\
ModelM2     & $1.28$       & $0.42$        & $0.94^{***}$  & $0.09$      & $0.75^{***}$  & $0.28$        & $0.92$        & $27.54$   & $1.68^{***}$  & $1.70^{***}$  & $0.22$        \\
            & $(0.64)$     & $(0.35)$      & $(0.18)$      & $(0.12)$    & $(0.18)$      & $(0.13)$      & $(0.35)$      & $(76.01)$ & $(0.20)$      & $(0.19)$      & $(0.15)$      \\
PersonaP2   & $-0.00$      & $1.06^{*}$    & $0.44^{*}$    & $0.14$      & $0.61^{**}$   & $0.12$        & $0.00$        & $-1.11$   & $-0.31$       & $0.38$        & $0.53^{**}$   \\
            & $(0.73)$     & $(0.43)$      & $(0.22)$      & $(0.14)$    & $(0.21)$      & $(0.16)$      & $(0.41)$      & $(1.16)$  & $(0.22)$      & $(0.20)$      & $(0.18)$      \\
PersonaP3   & $0.28$       & $0.39$        & $0.62^{**}$   & $-0.05$     & $-0.13$       & $-0.30$       & $-0.17$       & $0.30$    & $0.02$        & $-0.16$       & $-0.16$       \\
            & $(0.75)$     & $(0.45)$      & $(0.22)$      & $(0.14)$    & $(0.23)$      & $(0.16)$      & $(0.42)$      & $(0.77)$  & $(0.21)$      & $(0.21)$      & $(0.19)$      \\
\hline
AIC                       & 154.37     & 410.39      & 1068.72     & 2021.76   & 1061.31     & 1721.41     & 414.36      & 97.28   & 1065.69     & 1130.48     & 1420.38     \\
BIC                       & 181.17     & 437.18      & 1095.52     & 2048.56   & 1088.11     & 1748.20     & 441.16      & 124.08  & 1092.49     & 1157.27     & 1447.17     \\
Log Likelihood            & -72.19     & -200.19     & -529.36     & -1005.88  & -525.66     & -855.70     & -202.18     & -43.64  & -527.85     & -560.24     & -705.19     \\
Num. obs.                 & 1571       & 1571        & 1571        & 1571      & 1571        & 1571        & 1571        & 1571    & 1571        & 1571        & 1571        \\
Num. groups: Qs     & 262        & 262         & 262         & 262       & 262         & 262         & 262         & 262     & 262         & 262         & 262         \\
Var: Question (Int.) & 49.30      & 45.48       & 2.12        & 1.65      & 3.74        & 5.03        & 26.33       & 0.00    & 2.57        & 2.27        & 3.20        \\
\hline
\multicolumn{12}{l}{\scriptsize{$^{***}p<0.001$; $^{**}p<0.01$; $^{*}p<0.05$}}
\end{tabular}
\label{table:coefficientsLogHolmM2}
\end{center}
\end{table}
\end{landscape}

\begin{landscape}
    \begin{table}
\caption{Logistic Regression Results with Holmes Multiple Testing Correction on Persona Effect (P2 = Survey Design Expert) Only}
\begin{center}
\begin{tabular}{l c c c c c c c c c c c}
\hline
 & Code 1 & Code 2 & Code 3 & Code 4 & Code 5 & Code 6 & Code 7 & Code 8 & Code 9 & Code 10 & Code 11 \\
\hline
(Intercept) & $9.23^{***}$ & $-9.29^{***}$ & $-3.57^{***}$ & $-0.29^{*}$ & $-3.60^{***}$ & $-0.71^{***}$ & $-8.06^{***}$ & $-31.99$  & $-3.61^{***}$ & $-3.60^{***}$ & $-2.30^{***}$ \\
            & $(1.27)$     & $(0.90)$      & $(0.27)$      & $(0.14)$    & $(0.31)$      & $(0.19)$      & $(0.92)$      & $(76.01)$ & $(0.28)$      & $(0.27)$      & $(0.22)$      \\
ModelM2     & $1.28^{*}$   & $0.42$        & $0.94^{***}$  & $0.09$      & $0.75^{***}$  & $0.28^{*}$    & $0.92^{**}$   & $27.54$   & $1.68^{***}$  & $1.70^{***}$  & $0.22$        \\
            & $(0.64)$     & $(0.35)$      & $(0.18)$      & $(0.12)$    & $(0.18)$      & $(0.13)$      & $(0.35)$      & $(76.01)$ & $(0.20)$      & $(0.19)$      & $(0.15)$      \\
PersonaP2   & $-0.00$      & $1.06$        & $0.44$        & $0.14$      & $0.61^{*}$    & $0.12$        & $0.00$        & $-1.11$   & $-0.31$       & $0.38$        & $0.53^{*}$    \\
            & $(0.73)$     & $(0.43)$      & $(0.22)$      & $(0.14)$    & $(0.21)$      & $(0.16)$      & $(0.41)$      & $(1.16)$  & $(0.22)$      & $(0.20)$      & $(0.18)$      \\
PersonaP3   & $0.28$       & $0.39$        & $0.62^{**}$   & $-0.05$     & $-0.13$       & $-0.30$       & $-0.17$       & $0.30$    & $0.02$        & $-0.16$       & $-0.16$       \\
            & $(0.75)$     & $(0.45)$      & $(0.22)$      & $(0.14)$    & $(0.23)$      & $(0.16)$      & $(0.42)$      & $(0.77)$  & $(0.21)$      & $(0.21)$      & $(0.19)$      \\
\hline
AIC                       & 154.37     & 410.39      & 1068.72     & 2021.76   & 1061.31     & 1721.41     & 414.36      & 97.28   & 1065.69     & 1130.48     & 1420.38     \\
BIC                       & 181.17     & 437.18      & 1095.52     & 2048.56   & 1088.11     & 1748.20     & 441.16      & 124.08  & 1092.49     & 1157.27     & 1447.17     \\
Log Likelihood            & -72.19     & -200.19     & -529.36     & -1005.88  & -525.66     & -855.70     & -202.18     & -43.64  & -527.85     & -560.24     & -705.19     \\
Num. obs.                 & 1571       & 1571        & 1571        & 1571      & 1571        & 1571        & 1571        & 1571    & 1571        & 1571        & 1571        \\
Num. groups: Qs     & 262        & 262         & 262         & 262       & 262         & 262         & 262         & 262     & 262         & 262         & 262         \\
Var: Question (Int.) & 49.30      & 45.48       & 2.12        & 1.65      & 3.74        & 5.03        & 26.33       & 0.00    & 2.57        & 2.27        & 3.20        \\
\hline
\multicolumn{12}{l}{\scriptsize{$^{***}p<0.001$; $^{**}p<0.01$; $^{*}p<0.05$}}
\end{tabular}
\label{table:coefficientsLogHolmP2}
\end{center}
\end{table}
\end{landscape}

\begin{landscape}
    \begin{table}
\caption{Logistic Regression Results with Holmes Multiple Testing Correction on Persona Effect (P3 = Linguist) Only}
\begin{center}
\begin{tabular}{l c c c c c c c c c c c}
\hline
 & Code 1 & Code 2 & Code 3 & Code 4 & Code 5 & Code 6 & Code 7 & Code 8 & Code 9 & Code 10 & Code 11 \\
\hline
(Intercept) & $9.23^{***}$ & $-9.29^{***}$ & $-3.57^{***}$ & $-0.29^{*}$ & $-3.60^{***}$ & $-0.71^{***}$ & $-8.06^{***}$ & $-31.99$  & $-3.61^{***}$ & $-3.60^{***}$ & $-2.30^{***}$ \\
            & $(1.27)$     & $(0.90)$      & $(0.27)$      & $(0.14)$    & $(0.31)$      & $(0.19)$      & $(0.92)$      & $(76.01)$ & $(0.28)$      & $(0.27)$      & $(0.22)$      \\
ModelM2     & $1.28^{*}$   & $0.42$        & $0.94^{***}$  & $0.09$      & $0.75^{***}$  & $0.28^{*}$    & $0.92^{**}$   & $27.54$   & $1.68^{***}$  & $1.70^{***}$  & $0.22$        \\
            & $(0.64)$     & $(0.35)$      & $(0.18)$      & $(0.12)$    & $(0.18)$      & $(0.13)$      & $(0.35)$      & $(76.01)$ & $(0.20)$      & $(0.19)$      & $(0.15)$      \\
PersonaP2   & $-0.00$      & $1.06^{*}$    & $0.44^{*}$    & $0.14$      & $0.61^{**}$   & $0.12$        & $0.00$        & $-1.11$   & $-0.31$       & $0.38$        & $0.53^{**}$   \\
            & $(0.73)$     & $(0.43)$      & $(0.22)$      & $(0.14)$    & $(0.21)$      & $(0.16)$      & $(0.41)$      & $(1.16)$  & $(0.22)$      & $(0.20)$      & $(0.18)$      \\
PersonaP3   & $0.28$       & $0.39$        & $0.62^{*}$    & $-0.05$     & $-0.13$       & $-0.30$       & $-0.17$       & $0.30$    & $0.02$        & $-0.16$       & $-0.16$       \\
            & $(0.75)$     & $(0.45)$      & $(0.22)$      & $(0.14)$    & $(0.23)$      & $(0.16)$      & $(0.42)$      & $(0.77)$  & $(0.21)$      & $(0.21)$      & $(0.19)$      \\
\hline
AIC                       & 154.37     & 410.39      & 1068.72     & 2021.76   & 1061.31     & 1721.41     & 414.36      & 97.28   & 1065.69     & 1130.48     & 1420.38     \\
BIC                       & 181.17     & 437.18      & 1095.52     & 2048.56   & 1088.11     & 1748.20     & 441.16      & 124.08  & 1092.49     & 1157.27     & 1447.17     \\
Log Likelihood            & -72.19     & -200.19     & -529.36     & -1005.88  & -525.66     & -855.70     & -202.18     & -43.64  & -527.85     & -560.24     & -705.19     \\
Num. obs.                 & 1571       & 1571        & 1571        & 1571      & 1571        & 1571        & 1571        & 1571    & 1571        & 1571        & 1571        \\
Num. groups: Qs     & 262        & 262         & 262         & 262       & 262         & 262         & 262         & 262     & 262         & 262         & 262         \\
Var: Question (Int.) & 49.30      & 45.48       & 2.12        & 1.65      & 3.74        & 5.03        & 26.33       & 0.00    & 2.57        & 2.27        & 3.20        \\
\hline
\multicolumn{12}{l}{\scriptsize{$^{***}p<0.001$; $^{**}p<0.01$; $^{*}p<0.05$}}
\end{tabular}
\label{table:coefficientsLogHolmP3}
\end{center}
\end{table}
\end{landscape}

\subsection{Statement Order Analysis}

\begin{table}[htbp]
\centering
\caption{Average Statement Placement by Code}
    \label{tab:AvgStmtPlacement}
    \begin{tabular}{|c|c|}
    \hline 
    Code & Average Statement Placement\\ \hline
        Code 1 & 2.503495 \\ \hline
        Code 2 & 3.162500 \\ \hline
        Code 3 & 2.591549 \\ \hline
        Code 4 & 3.647551 \\ \hline
        Code 5 & 2.387387 \\ \hline
        Code 6 & 2.809917 \\ \hline
        Code 7 & 3.982759 \\ \hline
        Code 8 & 3.111111 \\ \hline
        Code 9 & 4.16682 \\ \hline
        Code 10 & 3.749004 \\ \hline
        Code 11 & 3.643060 \\ \hline
        SysVar & 4.243108 \\ \hline
        NOTA & 3.767263 \\ \hline
    \end{tabular}
    
\end{table}

\begin{table}[htbp]
\centering
\caption{One Way ANOVA Test for Different Average Statement Placement of Codes}
\label{tab:stmtorderANOVA}
\begin{tabular}{lllllll}
\toprule
Effect & \multicolumn{1}{c}{$\hat{\eta}^2_G$} & \multicolumn{1}{c}{90\% CI} & \multicolumn{1}{c}{$F$} & \multicolumn{1}{c}{$\mathit{df}$} & \multicolumn{1}{c}{$\mathit{df}_{\mathrm{res}}$} & \multicolumn{1}{c}{$p$}\\
\midrule
Code & .197 & {}[.183, .208] & 165.95 & 12 & 8140 & $<$ .001\\
\bottomrule
\end{tabular}

\end{table}

\begin{longtable}[htbp]{>{}lrrrr}
\caption{Pairwise Comparisons of Average Code Placement with Confidence Intervals and Adjusted $p$-Values} \\
\toprule
 & Difference & Lower CI & Upper CI & Adjusted p-Value \\
\midrule
\endfirsthead

\toprule
 & Difference & Lower CI & Upper CI & Adjusted p-Value \\
\midrule
\endhead

\midrule
\multicolumn{5}{r}{\textit{(continued on next page)}} \\
\midrule
\endfoot

\bottomrule
\endlastfoot

\cellcolor[HTML]{79C9C6}{Code 10-Code 1} & 1.2455092 & 0.9723676 & 1.5186509 & 0.0000000\\
\cellcolor[HTML]{79C9C6}{Code 11-Code 1} & 1.1395647 & 0.9064987 & 1.3726308 & 0.0000000\\
\cellcolor[HTML]{79C9C6}{Code 2-Code 1} & 0.6590052 & 0.1850065 & 1.1330040 & 0.0003053\\
\cellcolor[HTML]{C1BBB5}{Code 3-Code 1} & 0.0880545 & -0.2071261 & 0.3832352 & 0.9989028\\
\cellcolor[HTML]{79C9C6}{Code 4-Code 1} & 1.1440560 & 0.9845168 & 1.3035952 & 0.0000000\\
\addlinespace
\cellcolor[HTML]{C1BBB5}{Code 5-Code 1} & -0.1161074 & -0.4055510 & 0.1733362 & 0.9833451\\
\cellcolor[HTML]{79C9C6}{Code 6-Code 1} & 0.3064226 & 0.1370953 & 0.4757499 & 0.0000002\\
\cellcolor[HTML]{79C9C6}{Code 7-Code 1} & 1.4792639 & 0.9240813 & 2.0344464 & 0.0000000\\
\cellcolor[HTML]{C1BBB5}{Code 8-Code 1} & 0.6076164 & -0.7932429 & 2.0084756 & 0.9686212\\
\cellcolor[HTML]{79C9C6}{Code 9-Code 1} & 1.6531873 & 1.3606012 & 1.9457733 & 0.0000000\\
\addlinespace
\cellcolor[HTML]{79C9C6}{NOTA-Code 1} & 1.2637687 & 1.0996543 & 1.4278831 & 0.0000000\\
\cellcolor[HTML]{79C9C6}{SysVar-Code 1} & 1.7396130 & 1.5192395 & 1.9599866 & 0.0000000\\
\cellcolor[HTML]{C1BBB5}{Code 11-Code 10} & -0.1059445 & -0.4525398 & 0.2406508 & 0.9986055\\
\cellcolor[HTML]{79C9C6}{Code 2-Code 10} & -0.5865040 & -1.1254690 & -0.0475390 & 0.0189930\\
\cellcolor[HTML]{79C9C6}{Code 3-Code 10} & -1.1574547 & -1.5485303 & -0.7663790 & 0.0000000\\
\addlinespace
\cellcolor[HTML]{C1BBB5}{Code 4-Code 10} & -0.1014532 & -0.4035480 & 0.2006416 & 0.9965661\\
\cellcolor[HTML]{79C9C6}{Code 5-Code 10} & -1.3616166 & -1.7483803 & -0.9748529 & 0.0000000\\
\cellcolor[HTML]{79C9C6}{Code 6-Code 10} & -0.9390866 & -1.2464631 & -0.6317102 & 0.0000000\\
\cellcolor[HTML]{C1BBB5}{Code 7-Code 10} & 0.2337546 & -0.3778304 & 0.8453396 & 0.9889652\\
\cellcolor[HTML]{C1BBB5}{Code 8-Code 10} & -0.6378929 & -2.0620470 & 0.7862613 & 0.9597872\\
\addlinespace
\cellcolor[HTML]{79C9C6}{Code 9-Code 10} & 0.4076780 & 0.0185571 & 0.7967990 & 0.0300627\\
\cellcolor[HTML]{C1BBB5}{NOTA-Code 10} & 0.0182594 & -0.2862764 & 0.3227953 & 1.0000000\\
\cellcolor[HTML]{79C9C6}{SysVar-Code 10} & 0.4941038 & 0.1559130 & 0.8322946 & 0.0000985\\
\cellcolor[HTML]{C1BBB5}{Code 2-Code 11} & -0.4805595 & -1.0003640 & 0.0392450 & 0.1038954\\
\cellcolor[HTML]{79C9C6}{Code 3-Code 11} & -1.0515102 & -1.4157266 & -0.6872938 & 0.0000000\\
\addlinespace
\cellcolor[HTML]{C1BBB5}{Code 4-Code 11} & 0.0044913 & -0.2619195 & 0.2709021 & 1.0000000\\
\cellcolor[HTML]{79C9C6}{Code 5-Code 11} & -1.2556721 & -1.6152546 & -0.8960896 & 0.0000000\\
\cellcolor[HTML]{79C9C6}{Code 6-Code 11} & -0.8331421 & -1.1055273 & -0.5607569 & 0.0000000\\
\cellcolor[HTML]{C1BBB5}{Code 7-Code 11} & 0.3396991 & -0.2550694 & 0.9344677 & 0.8001042\\
\cellcolor[HTML]{C1BBB5}{Code 8-Code 11} & -0.5319484 & -1.9489623 & 0.8850656 & 0.9905658\\
\addlinespace
\cellcolor[HTML]{79C9C6}{Code 9-Code 11} & 0.5136225 & 0.1515058 & 0.8757393 & 0.0001962\\
\cellcolor[HTML]{C1BBB5}{NOTA-Code 11} & 0.1242039 & -0.1449716 & 0.3933795 & 0.9497620\\
\cellcolor[HTML]{79C9C6}{SysVar-Code 11} & 0.6000483 & 0.2933131 & 0.9067834 & 0.0000000\\
\cellcolor[HTML]{79C9C6}{Code 3-Code 2} & -0.5709507 & -1.1214128 & -0.0204886 & 0.0337155\\
\cellcolor[HTML]{C1BBB5}{Code 4-Code 2} & 0.4850508 & -0.0062024 & 0.9763040 & 0.0568701\\
\addlinespace
\cellcolor[HTML]{79C9C6}{Code 5-Code 2} & -0.7751126 & -1.3225197 & -0.2277055 & 0.0002039\\
\cellcolor[HTML]{C1BBB5}{Code 6-Code 2} & -0.3525826 & -0.8471013 & 0.1419360 & 0.4709051\\
\cellcolor[HTML]{79C9C6}{Code 7-Code 2} & 0.8202586 & 0.0963079 & 1.5442093 & 0.0111135\\
\cellcolor[HTML]{C1BBB5}{Code 8-Code 2} & -0.0513889 & -1.5272898 & 1.4245120 & 1.0000000\\
\cellcolor[HTML]{79C9C6}{Code 9-Code 2} & 0.9941820 & 0.4451069 & 1.5432572 & 0.0000002\\
\addlinespace
\cellcolor[HTML]{79C9C6}{NOTA-Code 2} & 0.6047634 & 0.1120054 & 1.0975215 & 0.0032882\\
\cellcolor[HTML]{79C9C6}{SysVar-Code 2} & 1.0806078 & 0.5663691 & 1.5948465 & 0.0000000\\
\cellcolor[HTML]{79C9C6}{Code 4-Code 3} & 1.0560015 & 0.7338423 & 1.3781607 & 0.0000000\\
\cellcolor[HTML]{C1BBB5}{Code 5-Code 3} & -0.2041619 & -0.6067925 & 0.1984687 & 0.9028244\\
\cellcolor[HTML]{C1BBB5}{Code 6-Code 3} & 0.2183681 & -0.1087490 & 0.5454851 & 0.5822618\\
\addlinespace
\cellcolor[HTML]{79C9C6}{Code 7-Code 3} & 1.3912093 & 0.7694686 & 2.0129500 & 0.0000000\\
\cellcolor[HTML]{C1BBB5}{Code 8-Code 3} & 0.5195618 & -0.9089830 & 1.9481066 & 0.9928737\\
\cellcolor[HTML]{79C9C6}{Code 9-Code 3} & 1.5651327 & 1.1602372 & 1.9700283 & 0.0000000\\
\cellcolor[HTML]{79C9C6}{NOTA-Code 3} & 1.1757141 & 0.8512649 & 1.5001634 & 0.0000000\\
\cellcolor[HTML]{79C9C6}{SysVar-Code 3} & 1.6515585 & 1.2953306 & 2.0077864 & 0.0000000\\
\addlinespace
\cellcolor[HTML]{79C9C6}{Code 5-Code 4} & -1.2601634 & -1.5770743 & -0.9432525 & 0.0000000\\
\cellcolor[HTML]{79C9C6}{Code 6-Code 4} & -0.8376334 & -1.0505340 & -0.6247328 & 0.0000000\\
\cellcolor[HTML]{C1BBB5}{Code 7-Code 4} & 0.3352078 & -0.2347769 & 0.9051926 & 0.7661595\\
\cellcolor[HTML]{C1BBB5}{Code 8-Code 4} & -0.5364397 & -1.9432309 & 0.8703515 & 0.9891860\\
\cellcolor[HTML]{79C9C6}{Code 9-Code 4} & 0.5091313 & 0.1893477 & 0.8289148 & 0.0000104\\
\addlinespace
\cellcolor[HTML]{C1BBB5}{NOTA-Code 4} & 0.1197127 & -0.0890658 & 0.3284911 & 0.7957456\\
\cellcolor[HTML]{79C9C6}{SysVar-Code 4} & 0.5955570 & 0.3401760 & 0.8509379 & 0.0000000\\
\cellcolor[HTML]{79C9C6}{Code 6-Code 5} & 0.4225300 & 0.1005804 & 0.7444795 & 0.0009907\\
\cellcolor[HTML]{79C9C6}{Code 7-Code 5} & 1.5953712 & 0.9763337 & 2.2144088 & 0.0000000\\
\cellcolor[HTML]{C1BBB5}{Code 8-Code 5} & 0.7237237 & -0.7036467 & 2.1510941 & 0.9028714\\
\addlinespace
\cellcolor[HTML]{79C9C6}{Code 9-Code 5} & 1.7692946 & 1.3685623 & 2.1700269 & 0.0000000\\
\cellcolor[HTML]{79C9C6}{NOTA-Code 5} & 1.3798760 & 1.0606374 & 1.6991147 & 0.0000000\\
\cellcolor[HTML]{79C9C6}{SysVar-Code 5} & 1.8557204 & 1.5042317 & 2.2072090 & 0.0000000\\
\cellcolor[HTML]{79C9C6}{Code 7-Code 6} & 1.1728413 & 0.6000397 & 1.7456428 & 0.0000000\\
\cellcolor[HTML]{C1BBB5}{Code 8-Code 6} & 0.3011938 & -1.1067411 & 1.7091286 & 0.9999658\\
\addlinespace
\cellcolor[HTML]{79C9C6}{Code 9-Code 6} & 1.3467647 & 1.0219870 & 1.6715423 & 0.0000000\\
\cellcolor[HTML]{79C9C6}{NOTA-Code 6} & 0.9573461 & 0.7409958 & 1.1736963 & 0.0000000\\
\cellcolor[HTML]{79C9C6}{SysVar-Code 6} & 1.4331904 & 1.1715831 & 1.6947978 & 0.0000000\\
\cellcolor[HTML]{C1BBB5}{Code 8-Code 7} & -0.8716475 & -2.3755878 & 0.6322928 & 0.7835397\\
\cellcolor[HTML]{C1BBB5}{Code 9-Code 7} & 0.1739234 & -0.4465897 & 0.7944365 & 0.9994116\\
\addlinespace
\cellcolor[HTML]{C1BBB5}{NOTA-Code 7} & -0.2154952 & -0.7867774 & 0.3557870 & 0.9901596\\
\cellcolor[HTML]{C1BBB5}{SysVar-Code 7} & 0.2603491 & -0.3295613 & 0.8502596 & 0.9640905\\
\cellcolor[HTML]{C1BBB5}{Code 9-Code 8} & 1.0455709 & -0.3824400 & 2.4735819 & 0.4252259\\
\cellcolor[HTML]{C1BBB5}{NOTA-Code 8} & 0.6561523 & -0.7511651 & 2.0634697 & 0.9457837\\
\cellcolor[HTML]{C1BBB5}{SysVar-Code 8} & 1.1319967 & -0.2829850 & 2.5469784 & 0.2799217\\
\addlinespace
\cellcolor[HTML]{79C9C6}{NOTA-Code 9} & -0.3894186 & -0.7115091 & -0.0673281 & 0.0042024\\
\cellcolor[HTML]{C1BBB5}{SysVar-Code 9} & 0.0864257 & -0.2676551 & 0.4405066 & 0.9998592\\
\cellcolor[HTML]{79C9C6}{SysVar-NOTA} & 0.4758443 & 0.2175805 & 0.7341082 & 0.0000001\\
\bottomrule
\end{longtable}

\subsection{Question Source Analysis}

\begin{table}[htbp]
\caption{Regression Results of Question Source Analysis}
\begin{center}
\begin{tabular}{l c}
\hline
 & Number of Codes \\
\hline
(Intercept)               & $2.30^{***}$ \\
                          & $(0.05)$     \\
ModelM2                   & $0.55^{***}$ \\
                          & $(0.04)$     \\
PersonaP2                 & $0.22^{***}$ \\
                          & $(0.05)$     \\
PersonaP3                 & $-0.03$      \\
                          & $(0.05)$     \\
SourceWVS                 & $0.19^{**}$  \\
                          & $(0.07)$     \\
SourceGallup              & $-0.31^{*}$  \\
                          & $(0.15)$     \\
\hline
AIC                       & $4114.94$    \\
BIC                       & $4157.82$    \\
Log Likelihood            & $-2049.47$   \\
Num. obs.                 & $1571$       \\
Num. groups: Qs     & $262$        \\
Var: Question (Int.) & $0.17$       \\
Var: Residual             & $0.67$       \\
\hline
\multicolumn{2}{l}{\scriptsize{$^{***}p<0.001$; $^{**}p<0.01$; $^{*}p<0.05$}}
\end{tabular}
\label{reg:Source}
\end{center}
\end{table}

\begin{landscape}

\begin{table}[htbp]
\caption{Question Source Regression Results with Holm's Multiple Testing Correction on SourceWVS}
\begin{center}
\begin{tabular}{l c c c c c c c c c c}
\hline
 & Code 1 & Code 2 & Code 3 & Code 4 & Code 5 & Code 6 & Code 7 & Code 8 & Code 9 & Code 10 \\
\hline
(Intercept)  & $0.98^{***}$ & $0.01$      & $0.07^{***}$ & $0.50^{***}$ & $0.10^{***}$ & $0.43^{***}$ & $0.02^{*}$ & $-0.00$     & $0.06^{**}$  & $0.02$       \\
             & $(0.01)$     & $(0.01)$    & $(0.02)$     & $(0.03)$     & $(0.02)$     & $(0.03)$     & $(0.01)$   & $(0.00)$    & $(0.02)$     & $(0.02)$     \\
ModelM2      & $0.01$       & $0.01$      & $0.08^{***}$ & $0.01$       & $0.06^{***}$ & $0.04^{*}$   & $0.02^{*}$ & $0.01^{**}$ & $0.14^{***}$ & $0.15^{***}$ \\
             & $(0.00)$     & $(0.01)$    & $(0.01)$     & $(0.02)$     & $(0.01)$     & $(0.02)$     & $(0.01)$   & $(0.00)$    & $(0.01)$     & $(0.02)$     \\
PersonaP2    & $-0.00$      & $0.02^{*}$  & $0.03$       & $0.03$       & $0.05^{**}$  & $0.02$       & $0.00$     & $-0.00$     & $-0.02$      & $0.04$       \\
             & $(0.01)$     & $(0.01)$    & $(0.02)$     & $(0.03)$     & $(0.02)$     & $(0.02)$     & $(0.01)$   & $(0.00)$    & $(0.02)$     & $(0.02)$     \\
PersonaP3    & $0.00$       & $0.01$      & $0.05^{**}$  & $-0.01$      & $-0.01$      & $-0.04$      & $-0.00$    & $0.00$      & $0.00$       & $-0.01$      \\
             & $(0.01)$     & $(0.01)$    & $(0.02)$     & $(0.03)$     & $(0.02)$     & $(0.02)$     & $(0.01)$   & $(0.00)$    & $(0.02)$     & $(0.02)$     \\
SourceWVS    & $-0.00$      & $0.07^{**}$ & $-0.03$      & $-0.13^{**}$ & $-0.03$      & $-0.12$      & $0.01$     & $0.00$      & $0.04$       & $0.15^{***}$ \\
             & $(0.01)$     & $(0.02)$    & $(0.02)$     & $(0.04)$     & $(0.03)$     & $(0.05)$     & $(0.02)$   & $(0.00)$    & $(0.03)$     & $(0.02)$     \\
SourceGallup & $0.01$       & $-0.01$     & $-0.08$      & $-0.24^{**}$ & $-0.03$      & $0.14$       & $0.02$     & $-0.00$     & $-0.03$      & $-0.02$      \\
             & $(0.02)$     & $(0.04)$    & $(0.05)$     & $(0.09)$     & $(0.06)$     & $(0.10)$     & $(0.03)$   & $(0.01)$    & $(0.06)$     & $(0.05)$     \\
\hline
\multicolumn{11}{l}{\scriptsize{$^{***}p<0.001$; $^{**}p<0.01$; $^{*}p<0.05$}}
\end{tabular}
\label{reg:Codes1to10SourceWVS}
\end{center}
\end{table}
\end{landscape}
\begin{landscape}
\begin{table}[htbp]
\caption{Question Source Regression Results with Holm's Multiple Testing Correction on SourceGallup}
\begin{center}
\begin{tabular}{l c c c c c c c c c c}
\hline
 & Code 1 & Code 2 & Code 3 & Code 4 & Code 5 & Code 6 & Code 7 & Code 8 & Code 9 & Code 10 \\
\hline
(Intercept)  & $0.98^{***}$ & $0.01$       & $0.07^{***}$ & $0.50^{***}$  & $0.10^{***}$ & $0.43^{***}$ & $0.02^{*}$ & $-0.00$     & $0.06^{**}$  & $0.02$       \\
             & $(0.01)$     & $(0.01)$     & $(0.02)$     & $(0.03)$      & $(0.02)$     & $(0.03)$     & $(0.01)$   & $(0.00)$    & $(0.02)$     & $(0.02)$     \\
ModelM2      & $0.01$       & $0.01$       & $0.08^{***}$ & $0.01$        & $0.06^{***}$ & $0.04^{*}$   & $0.02^{*}$ & $0.01^{**}$ & $0.14^{***}$ & $0.15^{***}$ \\
             & $(0.00)$     & $(0.01)$     & $(0.01)$     & $(0.02)$      & $(0.01)$     & $(0.02)$     & $(0.01)$   & $(0.00)$    & $(0.01)$     & $(0.02)$     \\
PersonaP2    & $-0.00$      & $0.02^{*}$   & $0.03$       & $0.03$        & $0.05^{**}$  & $0.02$       & $0.00$     & $-0.00$     & $-0.02$      & $0.04$       \\
             & $(0.01)$     & $(0.01)$     & $(0.02)$     & $(0.03)$      & $(0.02)$     & $(0.02)$     & $(0.01)$   & $(0.00)$    & $(0.02)$     & $(0.02)$     \\
PersonaP3    & $0.00$       & $0.01$       & $0.05^{**}$  & $-0.01$       & $-0.01$      & $-0.04$      & $-0.00$    & $0.00$      & $0.00$       & $-0.01$      \\
             & $(0.01)$     & $(0.01)$     & $(0.02)$     & $(0.03)$      & $(0.02)$     & $(0.02)$     & $(0.01)$   & $(0.00)$    & $(0.02)$     & $(0.02)$     \\
SourceWVS    & $-0.00$      & $0.07^{***}$ & $-0.03$      & $-0.13^{***}$ & $-0.03$      & $-0.12^{*}$  & $0.01$     & $0.00$      & $0.04$       & $0.15^{***}$ \\
             & $(0.01)$     & $(0.02)$     & $(0.02)$     & $(0.04)$      & $(0.03)$     & $(0.05)$     & $(0.02)$   & $(0.00)$    & $(0.03)$     & $(0.02)$     \\
SourceGallup & $0.01$       & $-0.01$      & $-0.08$      & $-0.24$       & $-0.03$      & $0.14$       & $0.02$     & $-0.00$     & $-0.03$      & $-0.02$      \\
             & $(0.02)$     & $(0.04)$     & $(0.05)$     & $(0.09)$      & $(0.06)$     & $(0.10)$     & $(0.03)$   & $(0.01)$    & $(0.06)$     & $(0.05)$     \\
\hline
\multicolumn{11}{l}{\scriptsize{$^{***}p<0.001$; $^{**}p<0.01$; $^{*}p<0.05$}}
\end{tabular}
\label{reg:Codes1to10SourceGallup}
\end{center}
\end{table}
\end{landscape}

\subsection{Comparison to Human Expert}

\begin{figure}[htbp]
    \centering
    \includegraphics[width=0.95\linewidth]{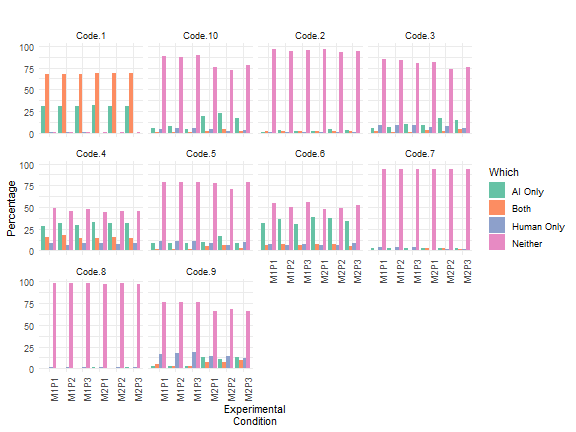}
    \caption{Comparison of Human and AI Use of Each Code}
    \label{fig:comparisontohuman}
\end{figure}

\end{appendices}

\newpage
\begin{singlespace}
\printbibliography
\end{singlespace}

\end{document}